\documentclass[onecolumn,epjc3,dvipsnames]{svjour3} %twocolumn?

\usepackage[space=true]{accsupp}
\usepackage{acronym}
\usepackage{amsmath}
\usepackage{amssymb}
\usepackage{babel}
\usepackage{booktabs}
\usepackage[utf8]{inputenc}
\usepackage[left=.09\paperwidth,right=.09\paperwidth,bottom=.1\paperheight,top=.1\paperheight]{geometry}
\usepackage{graphicx}
\graphicspath{{./figures/}}

\usepackage{imakeidx}
\makeindex[columns=3,title={Glossary of \EOS Classes \label{app:glossary}},options=-s glossary.ist]
\indexsetup{level=\section}

\usepackage[dvipsnames]{xcolor}
\definecolor{darkblue}{rgb}{0,0,0.9}
\usepackage{hyperref}
\hypersetup{
colorlinks,
citecolor=darkblue,
filecolor=darkblue,
linkcolor=darkblue,
urlcolor=darkblue
}

\usepackage{listings}
\usepackage[numbers,compress]{natbib}
\usepackage{newfloat}
\usepackage{placeins}
\usepackage{subfigure}
\usepackage[normalem]{ulem}
\usepackage{xcolor}
\usepackage{xspace}

\newcommand{\noncopynumber}[1]{%
    \BeginAccSupp{method=escape,ActualText={}}%
    #1%
    \EndAccSupp{}%
}
\newcommand{\copyablespace}{\BeginAccSupp{method=hex,unicode,ActualText=00A0}\ \EndAccSupp{}}

\definecolor{maroon}{cmyk}{0, 0.87, 0.68, 0.32}
\definecolor{halfgray}{gray}{0.55}
\definecolor{ipython_frame}{RGB}{207, 207, 207}
\definecolor{ipython_bg}{RGB}{247, 247, 247}
\definecolor{ipython_red}{RGB}{186, 33, 33}
\definecolor{ipython_green}{RGB}{0, 128, 0}
\definecolor{ipython_cyan}{RGB}{64, 128, 128}
\definecolor{ipython_purple}{RGB}{170, 34, 255}

\lstdefinelanguage{iPython}{
    morekeywords={access,and,break,class,continue,def,del,elif,else,except,exec,finally,for,from,global,if,import,in,is,lambda,not,or,pass,print,raise,return,try,while},%
    morekeywords=[2]{abs,all,any,basestring,bin,bool,bytearray,callable,chr,classmethod,cmp,compile,complex,delattr,dict,dir,divmod,enumerate,eval,execfile,file,filter,float,format,frozenset,getattr,globals,hasattr,hash,help,hex,id,input,int,isinstance,issubclass,iter,len,list,locals,long,map,max,memoryview,min,next,object,oct,open,ord,pow,property,range,raw_input,reduce,reload,repr,reversed,round,set,setattr,slice,sorted,staticmethod,str,sum,super,tuple,type,unichr,unicode,vars,xrange,zip,apply,buffer,coerce,intern},%
    sensitive=true,%
    morecomment=[l]\#,%
    morestring=[b]',%
    morestring=[b]",%
    morestring=[s]{'''}{'''},% used for documentation text (multi-line strings)
    morestring=[s]{"""}{"""},% added by Philipp Matthias Hahn
    morestring=[s]{r'}{'},% `raw' strings
    morestring=[s]{r"}{"},%
    morestring=[s]{r'''}{'''},%
    morestring=[s]{r"""}{"""},%
    morestring=[s]{u'}{'},% unicode strings
    morestring=[s]{u"}{"},%
    morestring=[s]{u'''}{'''},%
    morestring=[s]{u"""}{"""},%
    literate=
    {á}{{\'a}}1 {é}{{\'e}}1 {í}{{\'i}}1 {ó}{{\'o}}1 {ú}{{\'u}}1
    {Á}{{\'A}}1 {É}{{\'E}}1 {Í}{{\'I}}1 {Ó}{{\'O}}1 {Ú}{{\'U}}1
    {à}{{\`a}}1 {è}{{\`e}}1 {ì}{{\`i}}1 {ò}{{\`o}}1 {ù}{{\`u}}1
    {À}{{\`A}}1 {È}{{\'E}}1 {Ì}{{\`I}}1 {Ò}{{\`O}}1 {Ù}{{\`U}}1
    {ä}{{\"a}}1 {ë}{{\"e}}1 {ï}{{\"i}}1 {ö}{{\"o}}1 {ü}{{\"u}}1
    {Ä}{{\"A}}1 {Ë}{{\"E}}1 {Ï}{{\"I}}1 {Ö}{{\"O}}1 {Ü}{{\"U}}1
    {â}{{\^a}}1 {ê}{{\^e}}1 {î}{{\^i}}1 {ô}{{\^o}}1 {û}{{\^u}}1
    {Â}{{\^A}}1 {Ê}{{\^E}}1 {Î}{{\^I}}1 {Ô}{{\^O}}1 {Û}{{\^U}}1
    {œ}{{\oe}}1 {Œ}{{\OE}}1 {æ}{{\ae}}1 {Æ}{{\AE}}1 {ß}{{\ss}}1
    {ç}{{\c c}}1 {Ç}{{\c C}}1 {ø}{{\o}}1 {å}{{\r a}}1 {Å}{{\r A}}1
    {€}{{\EUR}}1 {£}{{\pounds}}1
    {^}{{{\color{ipython_purple}\^{}}}}1
    {=}{{{\color{ipython_purple}=}}}1
    {+}{{{\color{ipython_purple}+}}}1
    {*}{{{\color{ipython_purple}$^\ast$}}}1
    {/}{{{\color{ipython_purple}/}}}1
    {?}{{{\color{ipython_purple}?}}}1
    {+=}{{{+=}}}1
    {-=}{{{-=}}}1
    {*=}{{{$^\ast$=}}}1
    {/=}{{{/=}}}1
    {\ }{{\copyablespace}}1,
    identifierstyle=\color{black}\ttfamily,
    commentstyle=\color{ipython_cyan}\ttfamily,
    stringstyle=\color{ipython_red}\ttfamily,
    keepspaces=true,
    showspaces=true,%false,
    showstringspaces=false,
    rulecolor=\color{ipython_frame},
    frame=single,
    frameround={t}{t}{t}{t},
    framexleftmargin=6mm,
    numbers=none,
    numberstyle=\tiny\color{halfgray}\noncopynumber,
    backgroundcolor=\color{ipython_bg},
    basicstyle=\scriptsize\ttfamily,
    keywordstyle=\color{ipython_green}\ttfamily,
    aboveskip=\smallskipamount,
    belowskip=\smallskipamount
}
\definecolor{maroon}{cmyk}{0, 0.87, 0.68, 0.32}
\definecolor{halfgray}{gray}{0.55}
\definecolor{bash_frame}{RGB}{207, 207, 207}
\definecolor{bash_bg}{RGB}{247, 247, 247}
\definecolor{bash_red}{RGB}{186, 33, 33}
\definecolor{bash_green}{RGB}{0, 128, 0}
\definecolor{bash_cyan}{RGB}{64, 128, 128}
\definecolor{bash_purple}{RGB}{170, 34, 255}

\lstdefinelanguage{bash}[]{sh}%
  {morekeywords={alias,bg,bind,builtin,caller,command,compgen,compopt,%
      complete,coproc,curl,declare,disown,dirs,enable,fc,fg,help,%
      history,jobs,let,local,logout,mapfile,printf,pushd,popd,%
      readarray,select,set,suspend,shopt,source,times,type,typeset,%
      ulimit,unalias,wait},%
   otherkeywords={ [, ], [[, ]], \{, \} }%
  }%

\lstdefinelanguage{bash}{
    morekeywords={awk,break,case,cat,cd,continue,do,done,echo,elif,else,%
      env,esac,eval,exec,exit,export,expr,false,fi,for,function,getopts,%
      hash,history,if,in,kill,login,newgrp,nice,nohup,ps,pwd,read,%
      readonly,return,set,sed,shift,test,then,times,trap,true,type,%
      ulimit,umask,unset,until,wait,while},%
    morecomment=[l]\#,%
    morestring=[d]",%
    alsoletter={*"'0123456789.},%
    alsoother={\{\=\}},%
    literate={{=}{{{=}}}1},%
    literate={\$\{}{{{{\bfseries{}\$\{}}}}2,%
    otherkeywords={ [, ], \{, \} },%
    identifierstyle=\color{black}\ttfamily,
    commentstyle=\color{bash_cyan}\ttfamily,
    stringstyle=\color{bash_red}\ttfamily,
    keepspaces=true,
    showspaces=false,
    showstringspaces=false,
    rulecolor=\color{bash_frame},
    frame=single,
    frameround={t}{t}{t}{t},
    framexleftmargin=6mm,
    numbers=none,
    numberstyle=\tiny\color{halfgray},
    backgroundcolor=\color{bash_bg},
    basicstyle=\scriptsize\ttfamily,
    keywordstyle=\color{bash_green}\ttfamily,
    aboveskip=\smallskipamount,
    belowskip=\smallskipamount,
}[keywords,comments,strings]%
\definecolor{maroon}{cmyk}{0, 0.87, 0.68, 0.32}
\definecolor{halfgray}{gray}{0.55}
\definecolor{yaml_frame}{RGB}{207, 207, 207}
\definecolor{yaml_bg}{RGB}{247, 247, 247}
\definecolor{yaml_red}{RGB}{186, 33, 33}
\definecolor{yaml_green}{RGB}{0, 128, 0}
\definecolor{yaml_cyan}{RGB}{64, 128, 128}
\definecolor{yaml_purple}{RGB}{170, 34, 255}

\newcommand\YAMLcolonstyle{\color{red}\mdseries}
\newcommand\YAMLkeystyle{\color{black}\bfseries}
\newcommand\YAMLvaluestyle{\color{blue}\mdseries}
\lstdefinelanguage{yaml}{
    morekeywords={true,false,null,y,n},%
    morecomment=[l]\#,%
    morestring=[d]",%
    alsoletter={*"'0123456789.},%
    alsoother={\{\=\}},%
    literate={{=}{{{=}}}1},%
    literate={\$\{}{{{{\bfseries{}\$\{}}}}2,%
    identifierstyle=\color{black}\ttfamily,
    commentstyle=\color{yaml_cyan}\ttfamily,
    stringstyle=\color{yaml_red}\ttfamily,
    keepspaces=true,
    showspaces=false,
    showstringspaces=false,
    rulecolor=\color{yaml_frame},
    frame=single,
    frameround={t}{t}{t}{t},
    framexleftmargin=6mm,
    numbers=none,
    numberstyle=\tiny\color{halfgray},
    backgroundcolor=\color{yaml_bg},
    basicstyle=\scriptsize\ttfamily\YAMLkeystyle,
    keywordstyle=\color{yaml_green}\ttfamily,
    aboveskip=\smallskipamount,
    belowskip=\smallskipamount,
    moredelim=[l][\color{orange}]{\&},
    moredelim=[l][\color{magenta}]{*},
    moredelim=**[il][\YAMLcolonstyle{:}\YAMLvaluestyle\ ]{:\ },   % switch to value style at :
    moredelim=**[il][\YAMLkeystyle{,}]{,},                          % switch back to key style at ,
    morestring=[b]',
    morestring=[b]",
    literate =    {---}{{\ProcessThreeDashes}}3
                {|}{{\textcolor{red}\textbar}}1 
                {\ -\ }{{\mdseries\ -\ }}3,
}[keywords,comments,strings]%

\DeclareFloatingEnvironment[
    fileext=loo,
    listname={List of Outputs},
    name=Output,
    placement=tbhp,
]{joutput}

\newacro{BSM}[BSM]{Beyond the Standard Model}
\newcommand{\BSM}{\ac{BSM}\xspace}
\newacro{CKM}[CKM]{Cabibbo-Kobayashi-Maskawa}
\newcommand{\CKM}{\ac{CKM}\xspace}
\newacro{KDE}[KDE]{kernel density estimate}
\newcommand{\KDE}{\ac{KDE}\xspace}
\newacro{PDF}[PDF]{Probability Density Function}
\newcommand{\PDF}{\ac{PDF}\xspace}
\newacro{SM}[SM]{Standard Model}
\newcommand{\SM}{\ac{SM}\xspace}
\newacro{WET}[WET]{Weak Effective Theory}
\newcommand{\WET}{\ac{WET}\xspace}
\newcommand{\cpp}{\texttt{C++}\xspace}
\newcommand{\EOS}{\texttt{EOS}\xspace}
\newcommand{\FlavBit}{\texttt{FlavBit}\xspace}
\newcommand{\flavio}{\texttt{flavio}\xspace}
\newcommand{\Jupyter}{\texttt{Jupyter}\xspace}
\newcommand{\matplotlib}{\texttt{matplotlib}\xspace}
\newcommand{\NumPy}{\texttt{NumPy}\xspace}
\newcommand{\Python}{\texttt{Python}\xspace}
\newcommand{\pypmc}{\texttt{pypmc}\xspace}
\newcommand{\SciPy}{\texttt{SciPy}\xspace}
\newcommand{\SuperISO}{\texttt{SuperISO}\xspace}
\newcommand{\WCxf}{\texttt{WCxf}\xspace}
\newcommand{\wilson}{\texttt{wilson}\xspace}
\newcommand{\YAML}{\texttt{YAML}\xspace}
\newcommand{\code}[1]{\lstinline{#1}} % short inline code
\newcommand{\command}[1]{\textcolor{black}{\lstinline{#1}\xspace}}
\newcommand{\pyclass}[1]{\textcolor{black}{\lstinline{#1}\xspace}}
\newcommand{\class}[1]{\lstinline{#1}\index{#1}\xspace}
\newcommand{\method}[2]{\lstinline{#2}\index{#1!#2}\xspace}
\newcommand{\object}[1]{\lstinline{#1}\xspace}
\newcommand{\refapp}[1]{appendix~\ref{app:#1}}

\newcommand{\reffig}[1]{figure~\ref{fig:#1}}
\newcommand{\reflst}[1]{listing~\ref{lst:#1}}
\newcommand{\refout}[1]{output~\ref{out:#1}}
\newcommand{\refsec}[1]{section~\ref{sec:#1}}
\newcommand{\GeV}{\ensuremath{\mathrm{GeV}}}

\renewcommand{\Re}{\operatorname*{Re}}
\renewcommand{\Im}{\operatorname*{Im}}
\newcommand{\vecth}{\ensuremath{\vec\vartheta}}
\newcommand{\vecx}{\ensuremath{\vec{x}}}
\newcommand{\vecnu}{\ensuremath{\vec\nu}}
\newcommand{\doi}[1]{\href{http://dx.doi.org/#1}{#1}}
\newcommand{\hlred}[1]{\textcolor{red}{#1}}
\newcommand{\hlgreen}[1]{\textcolor{ForestGreen}{#1}}
\newcommand{\hlorange}[1]{\textcolor{orange}{#1}}
\makeatletter
    \def\ng{\@ifstar\@@ng\@ng}
    \newcommand{\@ng}[1]{\textcolor{ForestGreen}{[\textbf{NG:} #1]}}
    \newcommand{\@@ng}[1]{\textcolor{ForestGreen}{#1}}
    \def\dvd{\@ifstar\@@dvd\@dvd}
    \newcommand{\@dvd}[1]{\textcolor{purple}{[\textbf{DvD:} #1]}}
    \newcommand{\@@dvd}[1]{\textcolor{purple}{#1}}
    \def\mr{\@ifstar\@@mr\@mr}
    \newcommand{\@mr}[1]{\textcolor{cyan}{[\textbf{MR:} #1]}}
    \newcommand{\@@mr}[1]{\textcolor{cyan}{#1}}
\makeatother

\journalname{Eur. Phys. J. C}

\begin{document}

\newcommand{\EOSversion}{1.0\xspace}

\title{%
    \includegraphics[width=0.15\linewidth]{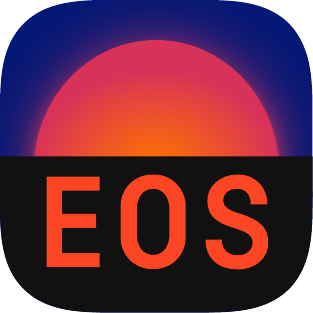}
    \hspace*{0.05\linewidth}
    \parbox[b][0.15\linewidth][c]{0.75\linewidth}{EOS\ \ ---\ \ A Software for Flavor Physics Phenomenology}\linebreak
}
\dedication{EOS-2021-04, TUM-HEP 1371/21, P3H-21-094, SI-HEP-2021-32}
\author{%
    The EOS Authors\thanksref{eMail} ---
    D.~van~Dyk\thanksref{aff:TUM} \and
    F.~Beaujean\thanksref{aff:C2PAP} \and
    T.~Blake\thanksref{aff:Warwick} \and
    C.~Bobeth\thanksref{aff:TUM} \and
    M.~Bordone\thanksref{aff:Torino} \and
    K.~Dugic\thanksref{aff:TUM} \and
    E.~Eberhard\thanksref{aff:TUM} \and
    N.~Gubernari\thanksref{aff:Siegen} \and
    E.~Graverini\thanksref{aff:EPFL} \and
    M.~Jung\thanksref{aff:Torino} \and
    A.~Kokulu \and
    S.~K\"urten\thanksref{aff:TUM} \and
    D.~Leljak\thanksref{aff:RBI} \and
    P.~L\"ughausen\thanksref{aff:TUM,aff:Origins} \and
    S.~Meiser\thanksref{aff:TUM} \and
    M.~Rahimi\thanksref{aff:Siegen} \and
    M.~Reboud\thanksref{aff:TUM} \and
    R.~Silva~Coutinho\thanksref{aff:UZH} \and
    J.~Virto\thanksref{aff:UB} \and
    K.K.~Vos\thanksref{aff:Maastricht,aff:NIKHEF}%
}
\institute{%
\setlength{\parindent}{0pt}
\label{aff:TUM}%
Physik Department T31, Technische Universit\"at M\"unchen, D-85748 Garching, Germany\and
\label{aff:C2PAP}
C2PAP, Excellence Cluster Universe, Ludwig-Maximilans-Universit\"at, D-85748 Garching, Germany\and
\label{aff:Warwick}
Department of Physics, University of Warwick, CV4\,7AL, Coventry, UK\and
\label{aff:Torino}
Dipartimento di Fisica, Universit\`a di Torino \& INFN, Sezione di Torino, I-10125 Torino, Italy\and
\label{aff:Siegen}
Theoretische Physik 1, Naturwissenschaftlich-Technische Fakult\"at, Universit\"at Siegen, D-57068 Siegen, Germany\and
\label{aff:EPFL}
Institute of Physics, \'{E}cole Polytechnique F\'{e}d\'{e}rale de Lausanne (EPFL), CH-1015 Lausanne, Switzerland
\and
\label{aff:RBI}
Rudjer Boskovic Institute, Division of Theoretical Physics, Bijeni\v cka 54, HR-10000 Zagreb, Croatia\and
\label{aff:Origins}
Excellence Cluster ORIGINS, Technische Universität München, D-85748 Garching, Germany\and
\label{aff:UZH}
Physik-Institut, Universit\"at Z\"urich, CH-8057 Z\"urich, Switzerland\and
\label{aff:UB}
Departament de Física Quàntica i Astrofísica and ICCUB,
Universitat de Barcelona, E08028 Barcelona, Catalunya\and
\label{aff:Maastricht}
Gravitational Waves and Fundamental Physics (GWFP),
Maastricht University, NL-6229 GT Maastricht, the Netherlands\and
\label{aff:NIKHEF}
Nikhef, NL-1098 XG Amsterdam, the Netherlands
}
\date{\today}
\thankstext{eMail}{\href{mailto:eos-developers@googlegroups.com}{eos-developers@googlegroups.com}}

\maketitle

\setlength{\parindent}{0pt}

\begin{abstract}
    \EOS is an open-source software for a variety of computational tasks in flavor physics.
    Its use cases include theory predictions within and beyond the Standard Model of particle physics,
    Bayesian inference of theory parameters from experimental and theoretical likelihoods, and
    simulation of pseudo events for a number of signal processes.
    \EOS ensures high-performance computations through a \cpp back-end and ease of usability through a \Python front-end.
    To achieve this flexibility, \EOS enables the user to select from a variety of implementations
    of the relevant decay processes and hadronic matrix elements at run time. In this article, we
    describe the general structure of the software framework and provide basic examples.
    Further details and in-depth interactive examples are provided as part of the \EOS online documentation.
\end{abstract}

\newpage

\setcounter{tocdepth}{3}
\renewcommand{\appendixname}{}
\tableofcontents

%
%
%--------+---------+---------+---------+---------+---------+---------+---------+
%
%
%--------+---------+---------+---------+---------+---------+---------+---------+
\section{Introduction and Physics Case}
\label{sec:intro}

Flavor physics phenomenology has a long history of substantial impact on the development of the \SM of particle physics.
Over the last decades, two developments are particularly noteworthy:\\
First, the determination of the \CKM matrix elements has developed into a precision enterprise,
thanks in large parts to the efforts at the B-factory experiments BaBar and Belle~\cite{Bevan:2014iga}
and more recently the LHCb experiment~\cite{LHCb:2012myk},
the technological progress in lattice gauge theory predictions~\cite{FlavourLatticeAveragingGroup:2019iem},
and the development of precision phenomenology with continuum methods.\\
Second, the emergence of the so-called ``$b$ anomalies'' has led to cautious excitement in the community.
These anomalies are substantial tensions
between theory predictions of $b$-quark decay observables and their measurements by the ATLAS, BaBar, Belle, CMS, and
LHCb experiments, which present a coherent pattern that might be due to \BSM effects,
but do not yet reach individually the required significance of $5\,\sigma$;
see e.g.\ refs.~\cite{Albrecht:2021tul,Bernlochner:2021vlv} for recent reviews.\\
Both developments have led to increasingly sophisticated phenomenological analyses.\\

Such analyses involve researchers regularly carrying out structurally similar and recurring tasks.
These typical \emph{use cases} include
\begin{enumerate}
    \item predicting flavor observables and assessing their theory uncertainties both within the \SM and for general
    \BSM scenarios in the \WET;
    \smallskip
    \item inferring hadronic, \SM, and/or \WET parameters from an extendable database of experimental and theoretical likelihoods;
    \smallskip
    \item simulating flavor processes and producing high-quality pseudo events for use in sensitivity studies and
    for the preparation of experimental analyses.
\end{enumerate}
The \EOS software~\cite{EOS} has been continuously developed since 2011~\cite{vanDyk:2012zla,EOS:repo} to achieve these tasks.
\EOS is free software published under the GNU General Public License 2~\cite{GPLv2}.
It has produced publication-quality results for approximately 30 peer-reviewed and published phenomenological studies~\cite{%
% 2010
Bobeth:2010wg,%
% 2011
Bobeth:2011gi,Bobeth:2011nj,%
% 2012
Beaujean:2012uj,Bobeth:2012vn,%
% 2013
Beaujean:2013soa,%
% 2014
Faller:2013dwa,SentitemsuImsong:2014plu,Boer:2014kda,%
% 2015
Beaujean:2015gba,Feldmann:2015xsa,Mannel:2015osa,%
% 2016
Bordone:2016tex,Meinel:2016grj,Boer:2016iez,Serra:2016ivr,%
% 2017
Bobeth:2017vxj,Blake:2017une,%
% 2018
Boer:2018vpx,Feldmann:2018kqr,Gubernari:2018wyi,%
% 2019
Boer:2019zmp,Bordone:2019vic,Blake:2019guk,Bordone:2019guc,%
% 2020
Gubernari:2020eft,%
% 2021
Bruggisser:2021duo,Leljak:2021vte,Bobeth:2021lya%
}.
Besides applications in phenomenology, \EOS also has been used in a number of published experimental studies
by the CDF~\cite{%
CDF:2011tds%
}, the CMS~\cite{%
CMS:2013mkz,%
CMS:2015bcy%
} and the LHCb~\cite{%
LHCb:2012bin,%
LHCb:2013zuf,%
LHCb:2014auh,%
LHCb:2015svh,%
LHCb:2018jna,%
LHCb:2020lmf%
} experiments.
The Belle II experiment has included \EOS as part of the external software~\cite{basf2ext} within the
Belle II software analysis framework~\cite{Kuhr:2018lps}.\\

In this article, we describe the \EOS user interface. Although the software is developed mainly in \cpp,
it is designed to be used in \Python \cite{python}.
As such, \EOS relies heavily
on the \code{numpy}~\cite{Harris:2020xlr} and \pypmc~\cite{pypmc} packages.
We \emph{highly} recommend new users to use \EOS within a \Jupyter notebook environment \cite{jupyter}.\\

\EOS can be installed in binary form on Linux-based systems\footnote{%
   This is limited to systems with \Python version 3.6 or later that also fulfill the ``manylinux\_2\_17\_x86\_64'' platform
   requirement as defined in PEP~600~\cite{PEP-600}.
}
with a single command:
\begin{lstlisting}[language=bash]
pip3 install eoshep
\end{lstlisting}
Afterwards, the \EOS \Python module can be accessed, e.g. within a \Jupyter notebook, using
\begin{lstlisting}[language=ipython]
import eos
\end{lstlisting}
We note that this means of installation also works for the ``Windows Subsystem for Linux v2 (WSL2)''. For the purpose
of installing \EOS, WSL2 can be treated like any Linux system.\\
Although \EOS can also be built and installed from source on macOS systems, we do not currently support these.
For macOS users we recommend to install on a remote-accessible Linux system and access a \Jupyter
notebook via \code{SSH}; our recommendation is described in detail as part of the frequently-asked questions~\cite{EOS:doc}.
Prospective \EOS developers find detailed instructions on how to build \EOS from source in the installation section of the documentation~\cite{EOS:doc}.\\

Presently, \EOS provides a total of 844 (pseudo-)observables\footnote{%
    \EOS does not distinguish between true observables, which can be unambiguously measured in an experimental
    setting, and pseudo-observables, which can not be unambiguously inferred from experimental or theoretical
    data. Pseudo-observables in \EOS include hadronic form factors and other auxiliary hadronic quantities.
} %
pertaining to a large variety of flavor processes.
Obtaining and browsing the full list of observables is discussed in \refsec{basics:observables}.
The processes implemented include
\begin{itemize}
    \item (semi)leptonic charged-current $\bar{B}$ meson decays (e.g., $\bar{B}\to D^*\tau\bar\nu$);
    \item semileptonic charged-current $\Lambda_b$ baryon decays (e.g., $\Lambda_b\to \Lambda_c(\to \Lambda \pi)\mu\bar\nu$);
    \item rare (semi)leptonic and radiative neutral-current $\bar{B}$ meson decays (e.g., $\bar{B}\to \bar{K}^*\mu^+\mu^-$);
    \item rare semileptonic and radiative neutral-current $\Lambda_b$ baryon decays (e.g., $\Lambda_b\to \Lambda(\to p \pi) \mu^+\mu^-$); and
    \item $B$-meson mixing observables (e.g., $\Delta m_s$).
\end{itemize}
\EOS is designed to be self documenting: a complete list of processes and their respective observables is automatically generated
as part of documentation, which is accessible both through the software itself and online~\cite{EOS:doc}.
The theoretical descriptions of most observables use the \WET to account for both \SM and \BSM predictions.
Details of the \EOS bases of \WET operators are described in \refapp{models:WET}.\\

Although \EOS is --- to our knowledge --- the first publicly available open-source flavor physics software~\cite{vanDyk:2012zla,EOS:repo}, 
it is by far not the only one. \EOS competes with the \flavio~\cite{Straub:2018kue}, \SuperISO~\cite{Neshatpour:2021nbn},
HEPfit~\cite{DeBlas:2019ehy} and \FlavBit~\cite{Workgroup:2017myk} software. Major distinctions between \EOS and these competitors are:
\begin{itemize}
    \item \EOS focuses on the simultaneous inference of hadronic and \BSM parameters;
    \item \EOS ensures modularity of hadronic matrix elements, i.e., the possibility to select from various hadronic models
    and parametrizations at run time;
    \item \EOS provides means to produce pseudo events for use in sensitivity studies and in preparation for experimental measurements; and
    \item \EOS provides means to predict hadronic matrix elements from QCD sum rules.
\end{itemize}
These distinctions make analyses possible that cannot currently be carried out with the competing software~\cite{%
Beaujean:2012uj,Beaujean:2013soa,Bobeth:2017vxj,Feldmann:2018kqr,bsll2021%
}, e.g., due to multi-modal or otherwise complicated posteriors that cannot be be captured by Markov chain Monte Carlo methods alone.
However, this benefit comes with an increased level of complexity, which we address in the \EOS documentation~\cite{EOS:doc}
and --- to some extent --- in this article.\\

\subsection{How to Read This Document}

Although this paper will give you a first impression of \EOS and basic examples to try in a \Jupyter notebook,
it is not meant to be a stand-alone document. To obtain a deeper understanding, additional documentation, and further
examples, the user is referred to refs.~\cite{EOS:doc,EOS:API,EOS:examples}.
Wherever we list \Python code, we assume that the reader evaluates it within a \Jupyter notebook environment,
to make full use its rich display capabilities.\\

In \refsec{basics}, we illustrate the basic
usage of \EOS, beginning with an overview of the various classes and concepts available
through the \Python interface.
In \refsec{usage} we continue with a discussion of and examples for the main use cases.
In a series of appendices we provide further details.
\begin{itemize}
    \item We describe the three physics models available in \EOS in \refapp{models}.
    \item We relegate lengthy \Python code examples that would otherwise interrupt reading \refsec{usage}
        to \refapp{PlotExamples}.
    \item We document the \EOS internal data format for storing experimental and theoretical likelihoods
        in \refapp{constraints-format}.
    \item We include a glossary of the main \EOS objects and associated methods in \refapp{glossary}.
\end{itemize}
This article is accompanied by a number of auxiliary files, containing example \Jupyter notebooks for
the basic usage and each of the use cases. These notebooks correspond to the examples
contained in the public source code repository~\cite{EOS:examples} as of \EOS version \EOSversion.

%
%
%--------+---------+---------+---------+---------+---------+---------+---------+
%
%
%--------+---------+---------+---------+---------+---------+---------+---------+
\section{Basic Classes and Concepts}
\label{sec:basics}

\EOS provides a number of \Python classes that make it possible to fulfill the physics use cases discussed in \refsec{usage}.
Three of the most relevant classes are used as follows:
\begin{itemize}
    \item hadronic and \BSM parameters are represented by objects of the \class{eos.Parameter} class;
    \smallskip
    \item physical observables and pseudo-observables (such as hadronic form factors) are represented by objects of the \class{eos.Observable} class;
    \smallskip
    \item likelihood functions, stemming from either experimental measurements or theoretical calculations, are represented by objects of the \class{eos.Constraint} class.
\end{itemize}
To facilitate their handling, \EOS has databases for all known objects of these classes. The user can interactively
inspect these databases within a \Jupyter notebook in the following way:
\begin{lstlisting}[language=iPython]
display(eos.Parameters())   # only run one line at a time, since the output is lengthy
display(eos.Observables())
display(eos.Constraints())
\end{lstlisting}
\EOS provides a rich display for most classes, including the above, which is not shown here for brevity.\\

All three databases can be searched by name of the target object. \EOS uses the same naming scheme for all three databases,
which is enforced through use of the \class{eos.QualifiedName} class.
The naming scheme is
\begin{center}
    % use texttt here instead of code for the color macros to be interpreted
    \texttt{\hlred{PREFIX}::\hlred{NAME}[@\hlred{SUFFIX}][;\hlred{OPTIONLIST}]}
\end{center}
where parts shown in square brackets are optional.
The individual parts have the following meaning:
\begin{description}
    \item[\texttt{\hlred{PREFIX}}] The prefix part is used to separate objects with (otherwise) identical names
    into different namespaces, to avoid conflicts. Examples of prefixes include
    parameter categories (e.g., \code{mass} or \code{decay-constant}), physical processes (e.g., \code{B->Kll}),
    or sectors of the \WET (e.g., \code{sbsb}).
    \smallskip
    
    \item[\texttt{\hlred{NAME}}] The name part is used to identify objects within its \code{PREFIX} namespace. Examples include observable names (e.g., \code{BR} for a branching ratio) or names of \WET Wilson coefficients (e.g., \code{cVL} for a coefficient of a left-handed
    vector operator).
    \smallskip
    
    \item[\texttt{\hlred{SUFFIX}}] The (optional) suffix part is used to distinguish between objects
    of otherwise identical names based on context. One example is
    the parameter describing $\Lambda_b$ baryon polarization, which takes different values based on the experimental
    environment. Generally, $\Lambda_b$ polarization would be represented by \code{Lambda_b::polarization}.
    The use of \code{@LHCb} and \code{@unpolarized} as a suffix distinguishes between the average polarization
    encountered within the LHCb experiment and an unpolarized setting (e.g. when using the whole phase space of the ATLAS and CMS experiments).
    \smallskip
    
    \item[\texttt{\hlred{OPTIONLIST}}] The option list is an optional comma-separated list of key/value pairs, which
    allows to modify the named object in an unambigious way. One example is \code{model=SM,l=mu,q=s}, which instructs
    an observable to use the Standard Model, $\mu$ lepton flavor, and strange-flavored spectator quarks.
    Details on possible options are discussed in \refsec{basics:observables}.
\end{description}

\begin{figure}[t]
    %\sidecaption
    \centering
    \includegraphics[width=.6\textwidth,trim=0 100 0 50,clip]{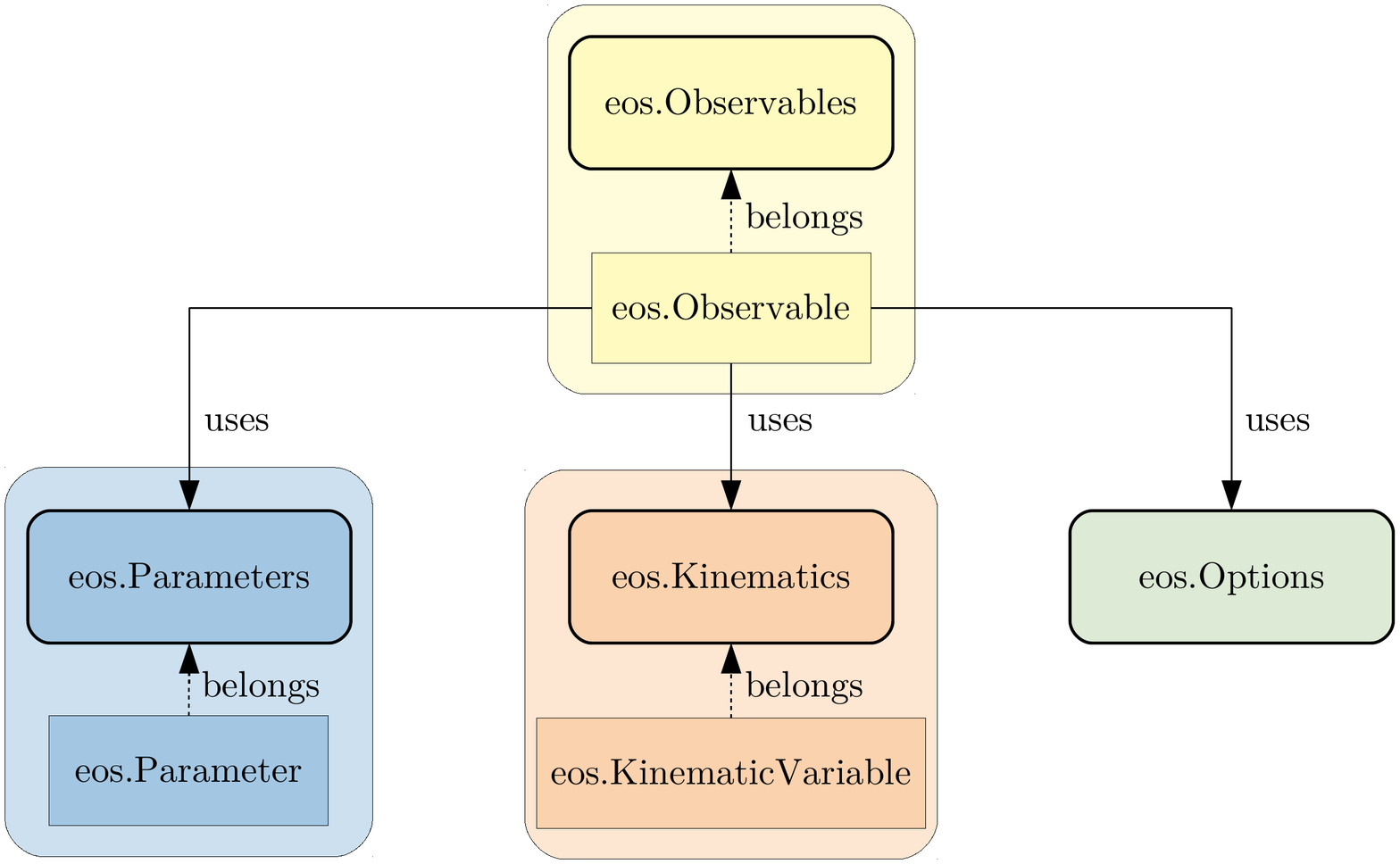}
    \caption{%
        Visual representation of the basic \EOS classes and their relationships.
        \label{fig:basics:class-diagram}
    }
\end{figure}

In the remainder of this section we discuss how to use the six representation
classes and their corresponding database classes
\begin{itemize}
    \item \class{eos.Parameter} within \class{eos.Parameters},
    \item \class{eos.KinematicVariable} within \class{eos.Kinematics},
    \item \class{eos.Option} within \class{eos.Options},
    \item \class{eos.Observable} within \class{eos.Observables},
    \item \class{eos.Constraint} within \class{eos.Constraints}, and
    \item \class{eos.SignalPDF} within \class{eos.SignalPDFs},
\end{itemize}
and the utility classes \class{eos.Analysis} and \class{eos.Plotter}. The relationship between the first four sets of classes
are illustrated in \reffig{basics:class-diagram}.
We provide a few examples here. However, for more exhaustive and interactive examples we refer to the notebook named
\href{https://github.com/eos/eos/tree/v1.0/examples/basics.ipynb}{basics.ipynb}, which is part of the collection of
\EOS example notebooks~\cite{EOS:examples}.

%
%
%
%--------+---------+---------+---------+---------+---------+---------+---------+
\subsection[Classes eos.Parameters and eos.Parameter]
{Classes \class{eos.Parameters} and \class{eos.Parameter}}
\label{sec:basics:parameters}

\EOS makes extensive use of the \class{eos.Parameter} class, which provides access
to a single real-valued scalar parameter. Any such \class{eos.Parameter} object is part of a large set of
built-in parameters. Users cannot directly create new objects of the \class{eos.Parameter} class.
However, new named sets of parameters can be created from which the parameter of interest can be extracted,
inspected, and altered.\\

We begin by creating and displaying a new set of parameters:
\begin{lstlisting}[language=iPython]
parameters = eos.Parameters()
display(parameters)
\end{lstlisting}
The new variable \object{parameters} now contains a representation of all parameters known to \EOS.
The \Jupyter \command{display} command has been augmented to provide a sectioned list of the known parameters,
which is rather lengthy and not shown here. It is
equivalent to the section ``List of Parameters`` in the \EOS documentation~\cite{EOS:doc}.
The display provides the user with an overview of all parameter names, their canonical physical
notation, and their value and unit.
A single parameter, here the muon mass as an example, can be isolated:
\begin{lstlisting}[language=iPython]
parameters['mass::mu']
\end{lstlisting}
Again, the user is provided with an overview of the parameter, including its qualified name,
unit, default value, and current value.
The value of an \class{eos.Parameter} object can be altered with the \method{eos.Parameter}{set} method:
\begin{lstlisting}[language=iPython]
m_mu = parameters['mass::mu']
display(m_mu)    # shows a value of 0.10566
m_mu.set(1.779)  # we just made the muon as heavy as the tauon!
display(m_mu)    # shows a value of 1.779
\end{lstlisting}
In this example, the muon mass parameter within \object{parameters} has been set to the measurd value for the tauon mass,
and the \object{m\_mu} object, which represents this parameter, has transparently changed its value.
Put differently: any \class{eos.Parameter} object ``remembers`` the set of parameters (i.e., the \class{eos.Parameters}
object) that it belongs to and forwards all changes to that set. To obtain
an independent set of parameters, the user can use
\begin{lstlisting}[language=iPython]
parameters2 = eos:Parameters()
parameters2['mass::mu']
display(parameters2 == parameters) # prints 'False', since the two sets are not identical!
\end{lstlisting}
A parameter's properties can be readily accessed through the methods \method{eos.Parameter}{name},
\method{eos.Parameter}{latex}, and \method{eos.Parameter}{evaluate}
\begin{lstlisting}[language=iPython]
display(m_mu.name())     # shows 'mass::mu'
display(m_mu.latex())    # shows 'm_\mu'
display(m_mu.evaluate()) # shows 1.779, since we changed it above.
\end{lstlisting}
A parameter object can be used like any other \Python object, e.g., as an element of a
\pyclass{list}, a \pyclass{dict}, or a \pyclass{tuple}:
\begin{lstlisting}[language=iPython]
lepton_masses = [parameters['mass::' + l] for l in ['e', 'mu', 'tau']]
[display(p) for p in lepton_masses]
translation = {p.name(): p.latex() for p in lepton_masses}
display(translation)
\end{lstlisting}
These properties allow to bind a function (e.g., the functional expression
of an observable or a likelihood function) to an arbitrary number of parameters,
let the function evaluate these parameters in a computationally efficient way, and let the
user change these parameters at a whim. Parameter sets are meant to be \emph{shared}, i.e.,
a single set of parameters is meant to be used by any number of functions. 
The sharing of parameters across observables makes it possible for \EOS to consistently and efficiently
evaluate a large number of functions.

The default set of parameters is stored in \YAML files that are installed together
with the binary \EOS library and the \Python modules and scripts.
The default parameter set can be replaced. To do this, the user must set the
environment variable \code{EOS_HOME} to point to an accessible directory.
The \YAML files found within \code{EOS_HOME/parameters} will be used \emph{instead} of
the default set of parameter contained in the \EOS package. The class \class{eos.Parameters} facilitate creating such files,
through the \method{eos.Parameters}{dump} method, which writes the current set
of parameters to a \YAML file. Alternatively, to use mostly the default parameter set, but override
a subset of parameters in persistent way, the user can use the
\method{eos.Parameters}{override\_from\_file} method to load only a subset of parameters from
a given file.

%
%
%--------+---------+---------+---------+---------+---------+---------+---------+
\subsection[Classes eos.Kinematics and eos.KinematicVariable]
{Classes \class{eos.Kinematics} and \class{eos.KinematicVariable}}
\label{sec:basics:kinematics}

\EOS uses the \class{eos.Kinematics} class to store a set of real-valued scalar kinematic variables by name. Contrary
to the class \class{eos.Parameters}, there are neither default variables nor default values.
Instead, \class{eos.Kinematics} objects are empty by default. Moreover, their
variables are only defined within the scope of a single \class{eos.Observable} object:
two observables that do not share an \class{eos.Kinematics} object can use identically-named, independent
kinematic variables.
Therefore, the names of kinematic variables do not require any prefix, and are simply (short) strings.\\

An empty set of kinematic variables can be created by
\begin{lstlisting}[language=iPython]
kinematics = eos.Kinematics()
\end{lstlisting}
A new kinematic variable can be declared with the existing (empty) set by providing
a key/value pair to the \object{kinematics} object, e.g.\index{eos.Kinematics!declare}
\begin{lstlisting}[language=iPython]
k1 = kinematics.declare('q2',             1.0)   # 1 GeV^2
k2 = kinematics.declare('E_pi',           0.139) # 139 MeV, a pion at rest!
k3 = kinematics.declare('cos(theta_pi)', -1.0)   # negative values are OK!
\end{lstlisting}
In this example, we have also captured the newly created kinematic variables as objects \object{k1}, \object{k2},
and \object{k3} of class \class{eos.KinematicVariable} for latter use.
\EOS uses the following guidelines for names and units of kinematic variables:
\begin{itemize}
    \item using \code{'q2'}, \code{'p2'}, and so on for the squares of four momenta $q^\mu$, $p^\mu$;
    \smallskip
    \item using \code{'E\_pi'}, \code{'E\_gamma'}, and so on for the energies of states $\pi$, $\gamma$
    in the rest frame of a decaying particle;
    \smallskip
    \item using \code{'cos(theta\_pi)'} and similar for the cosine of a helicity angle $\theta_\pi$;
    \smallskip
    \item using natural units, i.e., expressing all momenta and energies as powers of $\GeV$.
\end{itemize}
The new \class{eos.KinematicVariable} objects are now collected within the \object{kinematics} object.
They can be collectively inspected using
\begin{lstlisting}[language=iPython]
display(kinematics)
\end{lstlisting}
In addition, the individual objects \object{k1}, \object{k2}, etc.~can also be inspected
\begin{lstlisting}[language=iPython]
display(k1)
display(k2)
\end{lstlisting}
To directly obtain an \code{eos.Kinematics} object pre-populated with the variables one needs,
a \Python \pyclass{dict} can be provided to the constructor:\index{eos.Kinematics}
\begin{lstlisting}[language=iPython]
kinematics = eos.Kinematics({ 'q2': 1.0, 'E_pi': 0.139, 'cos(theta_pi)': -1.0 })
\end{lstlisting}
To extract a previously declared kinematic variable from the \object{kinematics} object, the \class{eos.Kinematics}
provides access via the subscript operator \code{[...]}
\begin{lstlisting}[language=iPython]
k1 = kinematics['q2']
k1.set(16.0)
\end{lstlisting}
In the above, the \method{eos.KinematicVariable}{set} method is used to change the value of \object{k1}.\\

Kinematic variables and their naming usually pertain to only one observable, which
will be discussed below in \refsec{basics:observables}.
Therefore, when creating observables, the user should create only a single independent set of kinematic variables
per observable. Nevertheless, it is possible to create observables that have a common set of
kinematic variables. This makes it possible to investigate correlations among observables that
share a kinematic variable (e.g., LFU ratios such as $R_K$ as a function of the lower dilepton momentum cut-off).

%
%
%--------+---------+---------+---------+---------+---------+---------+---------+
\subsection[Class eos.Options]
{Class \class{eos.Options}}
\label{sec:basics:options}

\EOS uses objects of the \code{eos.Options} class to modify the behavior of observables at runtime.
A new and empty set of options is created as follows\index{eos.Options}
\begin{lstlisting}[language=iPython]
options = eos.Options()
\end{lstlisting}
This object is usually populated with individual options, which are key/value pairs of \pyclass{str} objects.
Typical keys and their respective values include:
\begin{description}
    \item[\hlred{\texttt{model}}] is used to change the behavior of the low-energy observables.
    As of \EOS version \EOSversion, it can take the values \code{SM}, \code{CKM}, \code{WET}.
    \smallskip
    
    When choosing \code{SM}, the observables are computed within the \SM, and the values of the \WET Wilson coefficients
    are computed from \SM parameters. \CKM matrix elements are computed within the Wolfenstein parametrization.
    Details, such as the relevant parameter names, are discussed in \refapp{models:CKM}.
    \smallskip
    
    When choosing \code{CKM}, the observables are computed with \SM values for the \WET Wilson coefficients. However, the
    \CKM matrix elements are not computed from the Wolfenstein parameters. Instead, each \CKM matrix elements is parametrized
    in terms of two parameters for its absolute value and complex argument. This choice makes fitting \CKM matrix elements
    possible.
    Details, such as the relevant parameter names, are discussed in \refapp{models:CKM}.
    \smallskip
    
    When choosing \code{WET}, the observables are computed with generic values for the \WET Wilson coefficients. The
    \CKM matrix elements are treated as in the \code{CKM} case. This choice makes fitting \WET Wilson coefficients possible.
    Details, such as the \EOS convention for the basis of \WET operators and the relevant parameter names,
    are discussed in \refapp{models:WET}.
    \smallskip
    
    \item[\hlred{\texttt{form-factors}}] is used to select from one of the available parametrizations of hadronic form factors
    that are pertinent to the process. Its values are process-specific. For true observables (e.g., a semileptonic
    branching ratio) a sensible default choice is always provided. For pseudo-observables (e.g., the hadronic form factors
    $f_+(q^2)$ in $B\to \pi$ transitions) the choice must be made by the user.
    \smallskip
    
    \item[\hlred{\texttt{l}}] is used to select the charged lepton flavor in processes with at least one charged lepton.
    Allowed values are generally \code{e}, \code{mu} and \code{tau}. Individual processes might restrict the set of
    allowed values further, e.g., when hadronic matrix elements relevant to semitauonic decays are either unknown or unimplemented.
    \smallskip
    
    \item[\hlred{\texttt{q}}] is used to select the spectator quark flavor.
    Allowed values are typically \code{u}, \code{d}, \code{s}, and \code{c}. Individual processes might restrict
    the set of allowed values further. Processes with \code{s} and \code{c} spectator quarks are typically
    accessible through explicit specification of the spectator quark flavor in the process name,
    e.g., \code{B_s->K^*lnu}.
\end{description}
Obtaining the full list of option keys pertaining to a specific observable and their allowed keys is discussed in
\refsec{basics:observables}.\\

Adding new options to an existing \object{options} object is achieved as follows\index{eos.Options!declare}
\begin{lstlisting}[language=iPython]
options.declare('model', 'CKM')
options.declare('form-factors', 'BSZ2015')
options.declare('l', 'mu')                 # Since we are all so "cautiously excited"!
options.declare('q', 's')
display(options)
\end{lstlisting}
Analogously to the kinematic variables, an \class{eos.Options} object can be created
pre-populated with the values one needs using a \Python \pyclass{dict}
\begin{lstlisting}[language=iPython]
options = eos.Options({
    'form-factors': 'BSZ2015', # Bharucha, Straub, Zwicky 2015
    'model': 'WET',
    'l': 'tau',
    'q': 's'
})
\end{lstlisting}

%
%
%--------+---------+---------+---------+---------+---------+---------+---------+
\subsection[Classes eos.Observables and eos.Observable]
{Classes \class{eos.Observables} and \class{eos.Observable}}
\label{sec:basics:observables}

\EOS uses the \class{eos.Observable} class to provide theory predictions for a variety of
flavor physics processes and their associated (pseudo-)observables.
The complete list of observables known to \EOS is available as part of the online documentation~\cite[List of Observables]{EOS:doc}
and interactively in a \Jupyter notebook via\index{eos.Observables}
\begin{lstlisting}[language=iPython]
eos.Observables()
\end{lstlisting}
Within this list, all observables are uniquely identified by an \class{eos.QualifiedName} object; see
the beginning of \refsec{basics} for information on how such a name is structured.
To ease recognition, the typically used mathematical symbol for each observable is shown next to its name.
To search within this list, keyword arguments for the prefix part, name part, or suffix part of
a qualified name will filter the output. For example, the following code displays only branching ratios (\code{BR}) in processes
involving a $B^\mp$ meson (\code{B_u})\index{eos.Observables}
\begin{lstlisting}[language=iPython]
eos.Observables(prefix='B_u', name='BR')
\end{lstlisting}
Amongst others, this command lists the observable \code{B_u->lnu::BR}, representing the branching
ratio of $B^\mp\to \ell^\mp\bar\nu$ decays.
As part of the output the user is notified that this particular observable requires
no kinematic variables. The user is also notified about the \class{eos.Options} keys recognized
by this observable, which include \code{model} and \code{l}.\\

To create a new \class{eos.Observable} object the user needs to
\begin{itemize}
    \item identify it by name;
    \item provide a set of parameters that can optionally be shared with other observables;
    \item provide a set of kinematic variables that can optionally be shared with other observables; and
    \item specify the relevant options.
\end{itemize}
Again, the branching ratio of $B^\mp \to \ell^\mp\bar\nu$ is used as an example, specifically for a $\tau$
in the final state.
The observable is created as an \object{eos.Observable} object as follows\index{eos.Observable!make}
\begin{lstlisting}[language=iPython]
observable1 = eos.Observable.make('B_u->lnu::BR',
                                  eos.Parameters(), eos.Kinematics(), eos.Options({'l': 'tau', 'model': 'WET'}))
\end{lstlisting}
Here \code{B\_u->lnu::BR} is the \class{eos.QualifiedName} for
this particular observable, and default parameters are provided when using \code{eos.Parameters()}.
This observable does not
require any kinematic variables, and therefore an empty \class{eos.Kinematics} object is
provided. Setting the \code{l} option to \code{tau} selects a $\tau$ final state.
Setting the \code{model} option to \code{WET} enables the user to evaluate
the observable in the \WET for arbitrary values of the Wilson coefficients;
see \refapp{models:WET} for details.\\

The \class{eos.Observable} class provides access to the name, parameters, kinematics, options, and current value
of an observable by means of the following methods
\index{eos.Observable!name}\index{eos.Observable!parameters}\index{eos.Observable!kinematics}\index{eos.Observable!options}\index{eos.Observable!evaluate}
\begin{lstlisting}[language=iPython]
display(observable1.name())       # shows 'B_q->lnu::BR'
observable1.parameters()          # accesses the parameters
observable1.kinematics()          # accesses the (emtpy) set of kinematic variables
display(observable1.options())    # shows the options used to create the observable
display(observable1.evaluate())   # shows the current value
\end{lstlisting}

Note that each observable is associated with one object of the class \class{eos.Parameters}.
To illustrate this feature, the above code is repeated to create a second observable \object{observable2}\index{eos.Observable!make}
\begin{lstlisting}[language=iPython]
observable2 = eos.Observable.make('B_u->lnu::BR',
                                  eos.Parameters(), eos.Kinematics(), eos.Options({'l': 'tau', 'model': 'WET'}))
\end{lstlisting}
Even though the two objects \object{observable1} and \object{observable2}
share the same name and options, their respective parameter sets are
independent, as can be checked as follows:\index{eos.Observable!parameters}
\begin{lstlisting}[language=iPython]
observable1.parameters() == observable2.parameters()  # yields False
\end{lstlisting}
To correlate any number of observables, it is necessary to create \emph{all of them} using the same
\object{eos.Parameters} object; this will be further discussed in \refsec{usage:inference}.
In the above, this is \emph{not the case}, since for the creation of each observable the call
to \code{eos.Parameters()} created a new, independent set of parameters as explained in \refsec{basics:parameters}.\\

In many cases, observables have a default set of options, e.g., the default choice
of hadronic form factors or the default choice of a \BSM model. In some cases,
it does not make sense to have a default choice. In such cases, an error will be shown through a \Python exception
if the user does not provide a valid option value. An example for this behavior are the
form factor pseudo-observables, e.g., \code{B->K^*::V(q2)}, which always require providing
a valid value for the option \code{form-factors}. This is achieved
by including the option as part of the \class{eos.QualifiedName}. In this case,
\code{B->K^*::V(q2);form-factors=BSZ2015} selects the form factor parametrization as used
by Bharucha, Straub, and Zwicky~\cite{Straub:2015ica} in 2015.
A full list of all option keys and their respective valid values is available
as part of the online documentation~\cite{EOS:doc} and by displaying \texttt{eos.Observables()} in
an interactive \Jupyter notebook.\\

Contrary to parameters and kinematic variables, modifying the \class{eos.Options} object of any observable after its creation has no effect\index{eos.Observable!options}
\begin{lstlisting}[language=iPython]
observable1.options().set('l', 'mu') # does not affect observable1
\end{lstlisting}
This design decision ensures high-performance evaluations of all observables.\\

Objects of type \class{eos.Observable} are regular \Python objects. For example, they can be collected in a \pyclass{list},
which is useful for evaluating a number of identical observables at different points in their phase space.
This can be achieved as follows\index{eos.Observable!make}\index{eos.Observable!evaluate}
\begin{lstlisting}[language=iPython]
import numpy
parameters  = eos.Parameters()
observables = [eos.Observable.make('B->D^*lnu::A_FB(q2)', parameters, eos.Kinematics(q2=q2_value), eos.Options())
               for q2_value in numpy.linspace(1.00, 10.67, 10)]
values      = [o.evaluate() for o in observables]
display(values)
\end{lstlisting}
Here the instantiation of all observables with the same \class{eos.Parameters} object \object{parameters} ensures that
they share the same numerical values for all parameters. As a consequence, changes of numerical values within \object{parameters}
are broadcasted to all these instances and are taken into account in their subsequent evaluations.

%
%
%--------+---------+---------+---------+---------+---------+---------+---------+
\subsection[Classes eos.Constraints and eos.Constraint]
{Classes \class{eos.Constraints} and \class{eos.Constraint}}
\label{sec:basics:constraints}

\EOS uses the class \class{eos.Constraint} to manage and create individual likelihood functions
at run time. To this end, objects of type \class{eos.Constraint} contain both
information on the concrete likelihood (e.g., mean values and standard deviation of a
Gaussian measurement)
and meta-information about the constrained observables (e.g., the \EOS internal
names for an observable, relevant kinematic variables, and required options).

Besides (multivariate) Gaussian likelihood functions, \EOS
also supports LogGamma and Amoroso functions~\cite{crooks2015amoroso}, and Gaussian mixture densities.
The database of constraints makes it possible to construct a likelihood function
for any experimental measurement and/or theory input in terms of \class{eos.Observable} objects.
Hence, \class{eos.Constraint} objects are the building blocks for
parameter inference studies that use the \EOS software.\\

\EOS provides a database of constraints, which is available as part of the 
online documentation~\cite[List of Constraints]{EOS:doc} as well as interactively accessible in a \Jupyter
notebook via\index{eos.Constraints}
\begin{lstlisting}[language=iPython]
display(eos.Constraints())
\end{lstlisting}
This database is stored within \EOS in a series of \YAML files.
Most \EOS users will not require knowledge about the file format. However, advanced users
may need to provide constraints that are not part of the built-in database. In such
a case, the user can specify a \emph{manual} constraint; see \refsec{basics:analysis} and ref.~\cite{EOS:API} for details.
Alternatively, similar to the \class{eos.Parameters} database, the user can set the \code{EOS_HOME}
environment variable to point to an accessible directory. All \YAML files within \code{EOS_HOME/constraints}
will be loaded and used \emph{instead} of the default \class{eos.Constraints} database.
We document the format in \refapp{constraints-format} and an example entry is shown
in \reflst{constraint:Belle2015A}.\\

Examples of built-in constraints that are used later on in this document include:
\begin{itemize}
    \item The constraint \code{B->D::f_++f_0@FNAL+MILC:2015B} describes a lattice QCD result for
    the $\bar{B}\to D$ form factors $f_+$ and $f_0$. Here the suffix 
    indicates that this constraint has been extracted from ref.~\cite{Lattice:2015rga},
    which is included in the \EOS list of references as \code{FNAL+MILC:2015B}.
    The constraint can be used to create a likelihood function for the model parameters
    for the form factors $f_+$ and $f_0$,
    e.g., when using the BSZ2015 parametrization as the form factor model. Using
    \code{B->D::f_++f_0@FNAL/MILC:2015B;form-factors=BSZ2015} (i.e., the constraint name including
    an option list that specificies the form factor model) ensures that the correct form factor model
    (here: BSZ2015) is used when creating a likelihood from this constraint.
    \vspace*{\smallskipamount}
    \item The constraint \code{B^0->D^+e^-nu::BRs@Belle:2015A} describes the correlated
    measurement of the $\bar{B}^0\to D^+e^-\bar\nu$ branching ratio in $10$ bins of the kinematic
    variable $q^2$. Here the suffix indicates that the results have been extracted from ref.~\cite{Belle:2015pkj},
    which is included in the \EOS list of references as \code{Belle:2015A}.
\end{itemize}

%
%
%--------+---------+---------+---------+---------+---------+---------+---------+
\subsection[Classes eos.SignalPDFs and eos.SignalPDF]
{Classes \class{eos.SignalPDFs} and \class{eos.SignalPDF}}
\label{sec:basics:signalpdf}

\EOS uses the \class{eos.SignalPDF} class to provide a theoretical prediction
for the \PDF that describes a physical process, be it
a decay or a scattering process. The dependence on an arbitrary number of
kinematic variables are modeled through a shared object of class \class{eos.Kinematics},
and its \class{eos.KinematicVariable} objects. Parameters can be modified
or inferred through a shared \class{eos.Parameters} object. Hence,
each \class{eos.SignalPDF} object works very similar to an \class{eos.Observable}
object.
The list of \acp{PDF} can be accessed using the \class{eos.SignalPDFs} class.
Searching for a specific \PDF in the \EOS database of signal PDFs is possible
by filtering by the prefix part, name part, or suffix part of the signal \PDF qualified name,
very similar to how the database of observables is searchable
\begin{lstlisting}[language=iPython]
eos.SignalPDFs(prefix='B->Dlnu') # display a list of all known SignalPDF objects
                                 # this include 'B->Dlnu::dGamma/dq2', which requires
                                 # 'q2_min', 'q2_max', 'q2' as kinematic variables.
\end{lstlisting}
The signal \PDF \code{B->Dlnu::dGamma/dq2} features one kinematic variable, \code{q2}.
Its boundaries are also passed by means of \class{eos.KinematicVariable} objects,
which are conventionally named \code{q2_min} and \code{q2_max}.
\begin{lstlisting}[language=iPython]
pdf = eos.SignalPDF.make('B->Dlnu::dGamma/dq2', eos.Parameters(),
                         eos.Kinematics({'q2_min': 0.0, 'q2_max': 10.0, 'q2': 5.0),
                         eos.Options({'l': 'mu', 'model': 'WET'}))
\end{lstlisting}
The \PDF's parameters, kinematics, and options can be accessed with eponymous methods.
This design permits the user some flexibility. It makes it possible to produce
pseudo-events within the \SM and in the generic \WET; see \refsec{usage:simulation}
for this use case. In addition, it enables unbinned likelihood fits; their description
goes beyond the scope of this document.

%
%
%--------+---------+---------+---------+---------+---------+---------+---------+
\subsection[Class eos.Analysis]
{Class \class{eos.Analysis}}
\label{sec:basics:analysis}

\EOS uses the \class{eos.Analysis} as an interface for the user to describe a Bayesian analysis
to infer one or more parameters.
When creating an \class{eos.Analysis} object, the following arguments are used:
\begin{description}
    \item[\hlred{\texttt{priors}}] is a mandatory \pyclass{list} describing the univariate priors.
    This argument must describe at least one prior.
    Each prior is described through a \pyclass{dict} object, the structure of which is documented
    as part of the \Python API documentation~\cite[\texttt{eos.Analysis}]{EOS:API}.
    \smallskip
    \item[\hlred{\texttt{likelihood}}] is a mandatory \pyclass{list} describing all the constraints
    that enter the likelihood. Each element is a \pyclass{str} or \class{eos.QualifiedName}, specifying
    a single constraint. Although it is a mandatory parameter, this list can be left empty.
    \smallskip
    \item[\hlred{\texttt{global\_options}}] is an optional \pyclass{dict} describing the options that
    will be applied to all the observables that enter the likelihood.
    Note that these global options \textit{override} those specified via the qualified name scheme.
    For example, in a \BSM analysis, it is useful to include \code{'model': 'WET'}
    as a global option, to ensure that all observables will be evaluated using a selectable point
    in the \WET parameter space.
    \smallskip
    \item[\hlred{\texttt{fixed\_parameters}}] is an optional \pyclass{dict} describing
    parameters that shall be fixed to non-default values as part of the analysis.
    For example, to carry out a \BSM analysis of $b\to c\tau\nu$ processes for a non-default renormalization
    scale, the user can set the scale parameter to a fixed value of $3\,\GeV$ using \code{'cbtaunutau::mu': '3.0'}.
    \smallskip
    \item[\hlred{\texttt{manual\_constraints}}] is an optional \pyclass{dict} describing constraints
    that are not yet included in the \EOS database of constraints. The constraint format is described in
    \refapp{constraints-format}.
    Note that to use any of the manual constraints as part of the likelihood, their qualified names
    must still be added to the \code{likelihood} argument. 
\end{description}

\enlargethispage{2em}
\begin{lstlisting}[
    language=iPython,%
    caption={%
        Example for a Bayesian analysis to extract the \CKM parameter $|V_{cb}|$ from
        $\bar{B}\to D\lbrace e^-,\mu^-\rbrace \bar\nu$ data by the Belle experiment
        and lattice QCD input by the HPQCD and Fermilab/MILC collaborations.
        \label{lst:basics:analysis:definition}\index{eos.Analysis}\index{eos.Parameter!set}
    }
]
analysis_args = {
  'global_options': { 'form-factors': 'BSZ2015', 'model': 'CKM' },
  'priors': [
    {'parameter': 'CKM::abs(V_cb)',           'min': 38e-3, 'max': 45e-3, 'type': 'uniform'},
    {'parameter': 'B->D::alpha^f+_0@BSZ2015', 'min':  0.0,  'max':  1.0,  'type': 'uniform'},
    {'parameter': 'B->D::alpha^f+_1@BSZ2015', 'min': -4.0,  'max': -1.0,  'type': 'uniform'},
    {'parameter': 'B->D::alpha^f+_2@BSZ2015', 'min':  4.0,  'max':  6.0,  'type': 'uniform'},
    {'parameter': 'B->D::alpha^f0_1@BSZ2015', 'min': -1.0,  'max':  2.0,  'type': 'uniform'},
    {'parameter': 'B->D::alpha^f0_2@BSZ2015', 'min': -2.0,  'max':  0.0,  'type': 'uniform'}
  ],
  'likelihood': [
    'B->D::f_++f_0@HPQCD:2015A',
    'B->D::f_++f_0@FNAL+MILC:2015B',
    'B^0->D^+e^-nu::BRs@Belle:2015A',
    'B^0->D^+mu^-nu::BRs@Belle:2015A'
  ]
}
analysis = eos.Analysis(**analysis_args)
\end{lstlisting}
In \reflst{basics:analysis:definition} we define a statistical analysis for the inference of $|V_{cb}|$
from measurements of the $\bar{B}\to D\ell^-\bar\nu$ branching ratios by the Belle experiment.
This example will be further discussed in \refsec{usage:analysis}.
First, we define all the arguments used in our analysis.
\begin{itemize}
    \item Using the \code{global_options}, we choose the \code{BSZ2015} parametrization~\cite{Straub:2015ica}
    to model the hadronic form factors that enter semileptonic $\bar{B}\to D$ transitions.
    We also choose the \code{CKM} model to ensure that $|V_{cb}|$ is represented by a single parameter.
    \smallskip
    \item Priors for both the $|V_{cb}|$ parameter and the \code{BSZ2015} parameters are
    described in \code{priors}. Here, each parameter is assigned a uniform prior, which is chosen to
    contain at least $98\%$ ($\sim 3\,\sigma$) of the \emph{ideal} posterior probability, i.e., the priors have been
    chosen to be wide enough to ``contain`` the posterior defined by this analysis.
    \smallskip
    \item The likelihood is defined through a list of constraints, which in the above includes both
    theoretical lattice QCD results as well as experimental measurements by the Belle collaboration.
    For the first part we combine the correlated lattice QCD results published by the Fermilab/MILC and HPQCD collaborations in 2015 \cite{Na:2015kha,MILC:2015uhg}.
    For the second part, we combine binned measurements of the branching ratio for
    $\bar{B}^0\to D^+e^-\bar\nu$ and $\bar{B}^0\to D^+\mu^-\bar\nu$ decay.
    We reiterate that \EOS treats genuine physical observables and pseudo-observables identically.
\end{itemize}

\noindent The class \class{eos.Analysis} further provides convenience methods to carry out the statistical analysis:
\begin{description}
    \item[\hlred{\texttt{optimize}}] \index{eos.Analysis!optimize} uses the \code{scipy.optimize} module to find the best
    fit point of the posterior. Optional parameters determine the abort condition for the optimization
    and the starting point.
    \smallskip
    \item[\hlred{\texttt{sample}}] \index{eos.Analysis!sample} uses the \code{pypmc} module to produce random variates
    of the posterior using an adaptive version of the Metropolis-Hastings algorithm~\cite{doi:10.1063/1.1699114,10.1093/biomet/57.1.97,10.2307/3318737}
    with a single Markov chain.
    This method can be run several times to repeatedly explore the posterior density and accurately sample from it.
    \smallskip
    \item[\hlred{\texttt{sample\_pmc}}] \index{eos.Analysis!sample\_pmc} uses the \code{pypmc} module to produce random
    variates of the posterior using the Population Monte Carlo algorithm~\cite{2010MNRAS.405.2381K}. To this end, an initial guess
    of the posterior in form of a Gaussian mixture density is created~\cite{2013arXiv1304.7808B} from Markov chain Monte Carlo
    samples obtained using \method{eos.Analysis}{sample}.
\end{description}
At any point, the attribute \method{eos.Analysis}{parameters} can be used to access the analysis' parameter set,
e.g., to save the set to file via the \method{eos.Parameters}{dump} method.
We refer to the documentation of the \EOS \Python API~\cite{EOS:API} for further information.\\

Note that the \cpp backend used by \class{eos.Analysis} parallelizes the evaluation of the likelihood function.
By default, the number of concurrent threads will match the number of available processors. Users who need to
limit this number (e.g., due to using \EOS on a multi-user system in parallel to other users' jobs) can do so by
setting the \texttt{EOS\_MAX\_THREADS} environment variable to the limit.

%
%
%--------+---------+---------+---------+---------+---------+---------+---------+
\subsection[Class eos.Plotter]
{Class \class{eos.Plotter}}
\label{sec:basics:plotter}

\EOS implements a versatile plotting framework based on the class \class{eos.Plotter},
which relies on \matplotlib~\cite{matplotlib} for the actual plotting.
Its input must be formatted as a dictionary containing two keys: 
\code{plot} contains metadata and \code{contents} describes the plot items.
The value associated to the \code{plot} key is a dictionary; it describes the layout of the plot, including axis labels,
positioning of the legend, and similar settings that affect the entire plot.
The value associated to the \code{contents} key is a list; it describes the contents of the plot, expressed in terms of
independent plot items. Possible types of plot items include points, bands, contours, histograms.

\begin{lstlisting}[
    language=iPython,%
    caption={%
        High-level description of the arguments for the \class{eos.Plotter} class.
        The plot will appear inline in a \Jupyter notebook (if \code{FILENAME} is not specified)
        or be written to \code{FILENAME} (if specified). In the latter case, the output format
        will be determined based on the file extension.
        \label{lst:basics:plotter:description}\index{eos.Plotter}
    }%
]
plot_desc = {
    'plot': {
        'x': { ... },       # description of the x axis
        'y': { ... },       # description of the y axis
        'legend': { ... },  # description of the legend
        ...                 # further layouting options
    },
    'contents': [
        { ... }, # first plot item
        { ... }, # second plot item
    ]
}
eos.plot.Plotter(plot_desc, FILENAME).plot()
\end{lstlisting}

Each of the items is represented by a dictionary that contains a \code{type} key and an optional \code{name} key.
A full description of all item types and their parameters is available as part of the \EOS \Python API documentation~\cite{EOS:API}.
Here, we provide a brief summary for the most common types, which are used within examples in the course of this document: 
\begin{description}
    \item[\hlred{\texttt{observable}}] \index{eos.Plotter!observable} plots a single \EOS observable without uncertainties as a function of one kinematic variable or one parameter.
    See \reflst{usage:BtoDlnu:BR} for an example.
    \smallskip
    \item[\hlred{\texttt{histogram}}] \index{eos.Plotter!histogram} 
    \item[\hlred{\texttt{histogram2D}}] \index{eos.Plotter!histogram2D} plots either a 1D or a 2D histogram of pre-existing random samples. These samples can be contained in \Python objects within the notebook's memory or contained in
    a datafile on disk.
    See \reflst{usage:plot-prior-prediction-int} and \reflst{simulation:histogram-2D} for examples.
    \smallskip
    \item[\hlred{\texttt{uncertainty}}] \index{eos.Plotter!uncertainty} plots the uncertainty band of an observables
    as a function of one kinematic variable or one parameter. The random samples for the observables
    can be contained in \Python objects within the notebook's memory or contained in
    a datafile on disk.
    See \reflst{usage:plot-prior-prediction-diff} for an example.
    \smallskip
    \item[\hlred{\texttt{constraint}}] \index{eos.Plotter!constraint} displays a constraint either from the \EOS library or a manually added constraint.
    See \reflst{inference:posterior_samples_uncertainties} for an example.
    \smallskip
\end{description}

Beyond \code{type} and \code{name} keys, all item types also recognise the following optional keys:
\begin{description}
    \item[\hlred{\texttt{alpha}}] \index{eos.Plotter!alpha} A \pyclass{float}, between 0.0 and 1.0, which describes
    the opacity of the plot item expressed as an alpha value.
    A value of 0.0 means completely transparent, 1.0 means completely opaque.
    \smallskip
    \item[\hlred{\texttt{color}}] \index{eos.Plotter!color} A \pyclass{str}, containing any valid \matplotlib color
    specification, which describes the color of the plot item. Defaults to one of the colors in the \matplotlib default color cycler.
    \smallskip
    \item[\hlred{\texttt{label}}] \index{eos.Plotter!label} A \pyclass{str}, containing LaTeX commands, which describes
    the label that appears in the plot’s legend for this plot item.
\end{description}
In \reflst{basics:plotter:description}, \texttt{FILENAME} is an optional argument naming the file into which the plot shall be placed.
The file format is automatically determined based on the file name extension.

%
%
%--------+---------+---------+---------+---------+---------+---------+---------+
\subsection[Classes eos.References, eos.Reference, and eos.ReferenceName]
{Classes \class{eos.References}, \class{eos.Reference}, and \class{eos.ReferenceName}}
\label{sec:basics:references}

\EOS strives to give complete credit to the various works that underpin the theory predictions and the experimental
and phenomenological analyses that provide likelihoods. To this end, \EOS keeps a database of bibliographical metadata,
which is accessible via the \class{eos.References} class. Each entry is a tuple of an \class{eos.ReferenceName} object
that uniquely identifies the reference and the metdata data of the reference as an \class{eos.Reference} object.
For a complete list of works used within \EOS, we refer to the documentation~\cite[List of References]{EOS:doc}.

Each observable provides a list of reference names, corresponding to the pertinent pieces of literature
that were used in their implementations. This list is obtained via the \method{eos.Observable}{references} method,
which returns a generator of \class{eos.ReferenceName} objects:
\begin{lstlisting}[language=iPython]
obs = eos.Observable.make('B_u->lnu::BR',
      eos.Parameters(), eos.Kinematics(), eos.Options({'l': 'tau', 'model': 'WET'}))
display([rn for rn in obs.references()]) # shows 'DBG:2013A', amongst others
\end{lstlisting}
Further information on this reference can be obtained from its \class{eos.Reference} object:
\begin{lstlisting}[language=iPython]
ref = eos.References()['DBG:2013A']
display(ref)                        # displays the reference's title, authors, and eprint hyperlink (if available)
\end{lstlisting}

In a similar way, by convention the suffix part of each \class{eos.Constraint} is a valid reference name.
Therefore, to look up the reference that provides a constraint (e.g., \code{B->D::f_++f_0@FNAL+MILC:2015B})
the user can look up the associated bibliographical metadata based on the name's suffix part:
\begin{lstlisting}[language=iPython]
display(eos.References()['FNAL+MILC:2015B'])
\end{lstlisting}

If you feel that your work should be listed as part of a reference for any of the \EOS observables, please
contact the authors to include it.

%
%
%--------+---------+---------+---------+---------+---------+---------+---------+
%
%
%--------+---------+---------+---------+---------+---------+---------+---------+
\section{Use Cases}
\label{sec:usage}

Each of the three major use cases introduced in \refsec{intro} is discussed in details
in sections \ref{sec:usage:predictions} to \ref{sec:usage:simulation}.

%
%
%--------+---------+---------+---------+---------+---------+---------+---------+
\subsection{Theory Predictions}
\label{sec:usage:predictions}

[\textit{The example developed in this section can be run interactively from the example notebook for theory predictions available
from ref.~\cite{EOS:repo}, file \href{https://github.com/eos/eos/tree/v1.0/examples/predictions.ipynb}{examples/predictions.ipynb}}]\\

\EOS is equipped to produce theory predictions including their parametric uncertainties
for any of its built-in observables using Bayesian statistics. This requires knowledge
of the probability density function (PDF) of the pertinent parameters.
Here and throughout we will denote the set of parameters as $\vecth$, with
\begin{equation*}
    \vecth \equiv (\vecx, \vecnu)\,
\end{equation*}
where $\vecx$ represents the parameters of interest, and
$\vecnu$ represents the nuisance parameters. This distinction is entirely a semantic one,
and no technical differences arise from treating a parameter either way.
Production of theory predictions then falls into one of the following cases:
\begin{enumerate}
    \item theory predictions for fixed values of all parameters $\vecth = \vecth^*$;
    \item \textit{a-priori} predictions with propagation of uncertainties due
          to the \emph{prior} PDF $P_0(\vecth)$;
    \item \textit{a-posteriori} predictions with propagation of uncertainties
          due to the \emph{posterior} PDF $P(\vecth|D)$, where $D$ represents
          some data $D$. 
\end{enumerate}

Case 1 has been already mentioned with the concluding example of
\refsec{basics:observables}. 
In \refsec{usage:predictions:fixed} we provide an example showcasing
how to efficiently obtain these predictions. Cases 2 and 3 can be handled identically
in a Monte-Carlo framework and are discussed collectively in \refsec{usage:predictions:sampling}.

%--------+---------+---------+---------+---------+---------+---------+---------+
\subsubsection{Direct Evaluation for Fixed Parameters}
\label{sec:usage:predictions:fixed}

In \refsec{basics} we have explained how to evaluate an observable for a single configuration of the kinematic variables,
e.g., an integrated branching ratio with fixed integration boundaries, or a differential branching ratio at one point
in the kinematic phase space. Commonly, users need to plot such differential observables as a function of the kinematic
variable but for fixed values of its parameters.
To illustrate how this can be achieved with \EOS, we use the differential branching ratios
for $\bar{B} \to D\lbrace \mu^-,\tau^-\rbrace \bar\nu$ as an example.
The \class{eos.Plotter} class (see \refsec{basics:plotter}), provides means to plot any
\EOS observable as a function of a single kinematic variable (here: $q^2$).
\begin{lstlisting}[%
    language=iPython,%
    caption={%
        Plot the $q^2$-differential branching ratios for $\bar{B} \to D\lbrace \mu^-,\tau^-\rbrace \bar\nu$.
        The results are shown as the two central curves in the right plot \refout{usage:prior-prediction}.
        \label{lst:usage:BtoDlnu:BR}\index{eos.Plotter!plot}
    }%
]
plot_args = {
    'plot': {
        'x': {'label': r'$q^2$', 'unit': r'$\textnormal{GeV}^2$', 'range': [0.0, 11.60] },
        'y': {'label': r'$d\mathcal{B}/dq^2$',                    'range': [0.0,  5e-3] },
        'legend': { 'location': 'upper center' }
    },
    'contents': [
        {
            'type': 'observable',  'observable': 'B->Dlnu::dBR/dq2;l=mu',
            'variable': 'q2',      'range': [0.02, 11.60],
            'label': r'$\ell=\mu$',
        },
        {
            'type': 'observable',  'observable': 'B->Dlnu::dBR/dq2;l=tau',
            'variable': 'q2',      'range': [3.17, 11.60],
            'label': r'$\ell=\tau$',
        }
    ]
}
eos.plot.Plotter(plot_args).plot()
\end{lstlisting}
The output is a plot containing the branching ratios for $\ell=\mu, \tau$,
where $x$ axis shows the kinematic variable $q^2$, and the $y$ axis shows the
value of the differential branching ratio. The output corresponds to the central curves
shown in the right plot of \refout{usage:prior-prediction}.
In the listing above, the statement \code{'variable': 'q2'} specifies that the
kinematic variable \code{q2} is varied in the available \code{range}.\\

Similarly, we can plot an observable as a function of a single parameter, with all other parameters
kept fixed and for a given kinematic configuration. To this end, the \code{'xrange'} requires
adjustment compared to the previous example, and the contents should be replaced by
\begin{lstlisting}[language=iPython]
...
    'contents': [
        {
            'type': 'observable',   'observable': 'B->Dlnu::dBR/dq2;l=mu,model=WET',
            'kinematics': {'q2': 2.0}, 'parameters': {'CKM::abs(V_cb)' : 0.042}
            'variable': 'cbmunumu::Re{cSL}', 'range': [-1.0, 1.0],
            'label': r'$\ell=\mu$',
        }
    ]
...
\end{lstlisting}
Here the dependence of the differential branching fraction at $q^2 = 2\,\GeV^2$
on the real part of the \WET Wilson coefficient $C_{S_L}$ in the $\bar{c}b\mu\nu_\mu$ sector of the \WET
is plotted. Note that \code{kinematics} key is used to provide the fixed set of kinematic
variables and the \code{parameters} key is used to modify parameter values.
As before, \code{variable} selects the entity that is plotted on the $x$ axis,
which is now recognized to be an \class{eos.Parameter} object rather than an \class{eos.KinematicVariable} object.

%--------+---------+---------+---------+---------+---------+---------+---------+
\subsubsection{Predictions from Monte Carlo Sampling}
\label{sec:usage:predictions:sampling}

\EOS provides the means for a more sophisticated estimation of theory uncertainties
using Monte Carlo techniques, including importance sampling techniques.
For the sampling of a probability density function, \EOS relies on the
\pypmc package that provides methods for adaptive
Metropolis-Hastings~\cite{doi:10.1063/1.1699114,10.1093/biomet/57.1.97,10.2307/3318737}
and Population Monte Carlo~\cite{2010MNRAS.405.2381K,2013arXiv1304.7808B} sampling.
The uncertainty of an observable $O$ is estimated
from its random variates. We recall that $O \sim P(O)$ with~\cite{gelmanbda04}
\begin{align}
    P(O) & 
    = \int\mathrm{d}\vecth\, P(O, \vecth)
    = \int\mathrm{d}\vecth\, P(O|\vecth) P(\vecth)
    = \int\mathrm{d}\vecth\,
       \delta\left[O - f_O(\vecth)\right] P(\vecth)\,.
\end{align}
Here the Dirac $\delta$-function was used and $f_O(\vecth)$ is the theoretical expression that
predicts $O$ for a given set of parameters \vecth. With this knowledge
at hand, we approach the two cases 2 and 3 as discussed in \ref{sec:usage:predictions}
in a basically identical way:\\

For \emph{case 2}, we use $P(\vecth) = P_0(\vecth)$, i.e., the prior PDF.
We note that \EOS treats all priors $P_0$ as \emph{univariate PDFs} and therefore as uncorrelated.
Mathematically, a multivariate prior is equivalent to a multivariate likelihood 
with flat, univariate priors.
By design, \EOS implements multivariate correlated priors in terms of a multivariate correlated
likelihood.
For example, the parameters in the parameterizations
of hadronic form factors are constrained by various theoretical methods like lattice
QCD calculations, light-cone sum rule calculations, unitarity bounds and
constraints that arise in the limit of a heavy-quark mass. Under these
circumstances one might still use the terminology \textit{prior prediction}
whenever the included constraints are only of theoretical nature, i.e. no
experimental information was used.\\

For \emph{case 3}, we use $P(\vecth) = P(\vecth | D)$, i.e., the posterior PDF
as obtained from a previous fit given some data $D$. Although based on case 3,
the examples below also illustrate case 2, since this distinction is entirely a semantic one.\\

We continue using the integrated branching ratios of $B^-\to D^0 \lbrace\mu^-,
\tau^-\rbrace\bar\nu$ decays as examples. The largest source of theoretical
uncertainty in these decays arises from the hadronic matrix elements, i.e.,
from the form factors $f^{\bar{B}\to D}_+(q^2)$ and $f^{\bar{B}\to D}_0(q^2)$.
Both form factors have been obtained independently using lattice QCD simulations
by the HPQCD \cite{Na:2015kha} and FNAL/MILC \cite{Lattice:2015rga} collaborations.
In the following this information is used as part of the data $D$
in the form of a joint likelihood. The form factors at different $q^2$ values
of each calculation are available in \EOS as \class{eos.Constraint} objects
under the names \code{B->D::f_++f_0@HPQCD:2015A} and
\code{B->D::f_++f_0@FNAL+MILC:2015B}.
Here, we use these two constraints to construct a multivariate Gaussian prior
as follows:
\begin{lstlisting}[language=iPython]
analysis_args = {
  'priors': [
    {'parameter': 'B->D::alpha^f+_0@BSZ2015', 'min': 0.0, 'max': 1.0, 'type': 'uniform'},
    {'parameter': 'B->D::alpha^f+_1@BSZ2015', 'min':-5.0, 'max':+5.0, 'type': 'uniform'},
    {'parameter': 'B->D::alpha^f+_2@BSZ2015', 'min':-5.0, 'max':+5.0, 'type': 'uniform'},
    {'parameter': 'B->D::alpha^f0_1@BSZ2015', 'min':-5.0, 'max':+5.0, 'type': 'uniform'},
    {'parameter': 'B->D::alpha^f0_2@BSZ2015', 'min':-5.0, 'max':+5.0, 'type': 'uniform'}
  ],
  'likelihood': [ 'B->D::f_++f_0@HPQCD:2015A', 'B->D::f_++f_0@FNAL+MILC:2015B' ]
}
prior = eos.Analysis(**analysis_args)
\end{lstlisting}

Next we create two observables: the semimuonic branching ratio and the
semitauonic branching ratio. By using
\object{prior.parameters} in the construction of these observables, we
ensure that our observables and the \object{prior} share the
same parameter set. This means that changes to \object{prior.parameters}
will affect the evaluation of both observables.
\begin{lstlisting}[language=iPython,
    caption={%
        Produce samples of the prior and prior-predictive samples for two observables.
        \label{lst:usage:prior-samples-int}
    }%
]
obs_mu  = eos.Observable.make('B->Dlnu::BR',  prior.parameters, eos.Kinematics({'q2_min': 0.02, 'q2_max': 11.60}),
                              eos.Options({'l':'mu', 'form-factors':'BSZ2015'}))
obs_tau = eos.Observable.make('B->Dlnu::BR',  prior.parameters, eos.Kinematics({'q2_min': 3.17, 'q2_max': 11.60}),
                              eos.Options({'l':'tau','form-factors':'BSZ2015'}))
observables = (obs_mu, obs_tau)
parameter_samples, _, observable_samples = prior.sample(N=5000, pre_N=1000, observables=observables)
\end{lstlisting}
In the above, we provide the option \code{'form-factors': 'BSZ2015'}
to ensure that the form factor plugin corresponds to the set of parameters
that are described by \object{prior}.
Sampling from the natural logarithm of the prior PDF and -- at the same time -- producing
prior-predictive samples of both observables is achieved using the \method{eos.Analysis}{sample} method.
This method runs one Markov chain using the \pypmc package, and it is discussed in more detail in
\refsec{usage:inference}.
Here \code{N=5000} samples of both the parameter set and the observable set are produced,
and we discard the values of the log prior for each parameter sample
by assigning the return value to \code{_}.
Note that the production of posterior-predictive samples is achieved in the same way. The distinction
between a prior PDF and a posterior PDF is entirely a semantic one.\\

To illustrate the prior-predictive samples we use \EOS' plotting framework:
\begin{lstlisting}[%
    language=iPython,%
    caption={%
        Histogram prior-predictive samples of two observables.
        The output is shown in the left plot \refout{usage:prior-prediction}.
        \label{lst:usage:plot-prior-prediction-int}\index{eos.Plotter!plot}
    }%
]
plot_args = {
    'plot': {
        'x': { 'label': r'$d\mathcal{B}/dq^2$',  'range': [0.0,  3e-2] },
        'legend': { 'location': 'upper center' }
    },
    'contents': [
        { 'label': r'$\ell=\mu$', 'type': 'histogram', 'bins': 30, 
          'data': { 'samples': observable_samples[:, 0] } },
        { 'label': r'$\ell=\tau$','type': 'histogram', 'bins': 30, 
          'data': { 'samples': observable_samples[:, 1] } },
    ]
}
eos.plot.Plotter(plot_args).plot()
\end{lstlisting}
The arithmetic mean and the variance of the samples can be determined with
standard techniques, e.g., using the \NumPy routines \code{numpy.average} and \code{numpy.var}.\\

A further recurring task is to produce and plot uncertainty bands for differential
observables. Here, we use the differential branching ratios for the previously discussed
semimuonic and semitauonic decays.
Using \EOS we approach this task by creating two lists of observables.
The first list includes only the $\bar{B}\to D\mu^-\bar\nu$ at various points in its phase space.
Due to the strong dependence of the branching ratio on $q^2$,
we do not distribute the points equally across the full phase space.
Instead, we equally distribute half of the points in the interval $[0.02\,\GeV^2, 1.00\,\GeV^2]$ and the other half in the remainder of the phase space.
The second list is constructed similarly for $\bar{B}\to D\tau^-\bar\nu$. We then pass these
lists to \method{eos.Analysis}{sample}, to obtain prior-predictive samples of the observables:
\begin{lstlisting}[%
    language=iPython,%
    caption={%
        Produce prior-predictive samples for the differential
        $\bar{B}\to D\lbrace \mu^-,\tau^-\rbrace \bar\nu$ branching ratios
        at various points in their respective phase spaces.
        The results are used in \reflst{usage:plot-prior-prediction-diff}
        to produce the output shown in the right plot of \refout{usage:prior-prediction}.
        \label{lst:usage:prior-samples-diff}
    }%
]
mu_q2values  = numpy.unique(numpy.concatenate((numpy.linspace(0.02,  1.00, 20), numpy.linspace(1.00, 11.60, 20))))
mu_obs       = [eos.Observable.make(
                   'B->Dlnu::dBR/dq2', prior.parameters, eos.Kinematics(q2=q2),
                   eos.Options({'form-factors': 'BSZ2015', 'l': 'mu'}))
               for q2 in mu_q2values]
tau_q2values = numpy.linspace(3.17, 11.60, 40)
tau_obs      = [eos.Observable.make(
                   'B->Dlnu::dBR/dq2', prior.parameters, eos.Kinematics(q2=q2),
                   eos.Options({'form-factors': 'BSZ2015', 'l': 'tau'}))
               for q2 in tau_q2values]
_, _, mu_samples  = prior.sample(N=5000, pre_N=1000, observables=mu_obs)
_, _, tau_samples = prior.sample(N=5000, pre_N=1000, observables=tau_obs)
\end{lstlisting}

We plot the so-obtained prior-predictive samples with \EOS's plotting framework:
\begin{lstlisting}[%
    language=iPython,%
    caption={%
        Plot the previously obtained prior-predictive samples.
        The production of the samples is achieved in \reflst{usage:prior-samples-diff},
        and the output is shown in the right plot of \refout{usage:prior-prediction}.
        \label{lst:usage:plot-prior-prediction-diff}\index{eos.Plotter!plot}
    }%
]
plot_args = {
    'plot': {
        'x': {'label': r'$q^2$', 'unit': r'$\textnormal{GeV}^2$', 'range': [0.0, 11.60] },
        'y': {'label': r'$d\mathcal{B}/dq^2$',                    'range': [0.0,  5e-3] },
        'legend': { 'location': 'upper center' }
    },
    'contents': [
        {
          'label': r'$\ell=\mu$', 'type': 'uncertainty', 'range': [0.02, 11.60],
          'data': { 'samples': mu_samples, 'xvalues': mu_q2values }
        },
        {
          'label': r'$\ell=\tau$','type': 'uncertainty', 'range': [3.17, 11.60],
          'data': { 'samples': tau_samples, 'xvalues': tau_q2values }
        },
    ]
}
eos.plot.Plotter(plot_args).plot()
\end{lstlisting}

\begin{joutput}[t]
    \centering
    \includegraphics[width=0.48\linewidth]{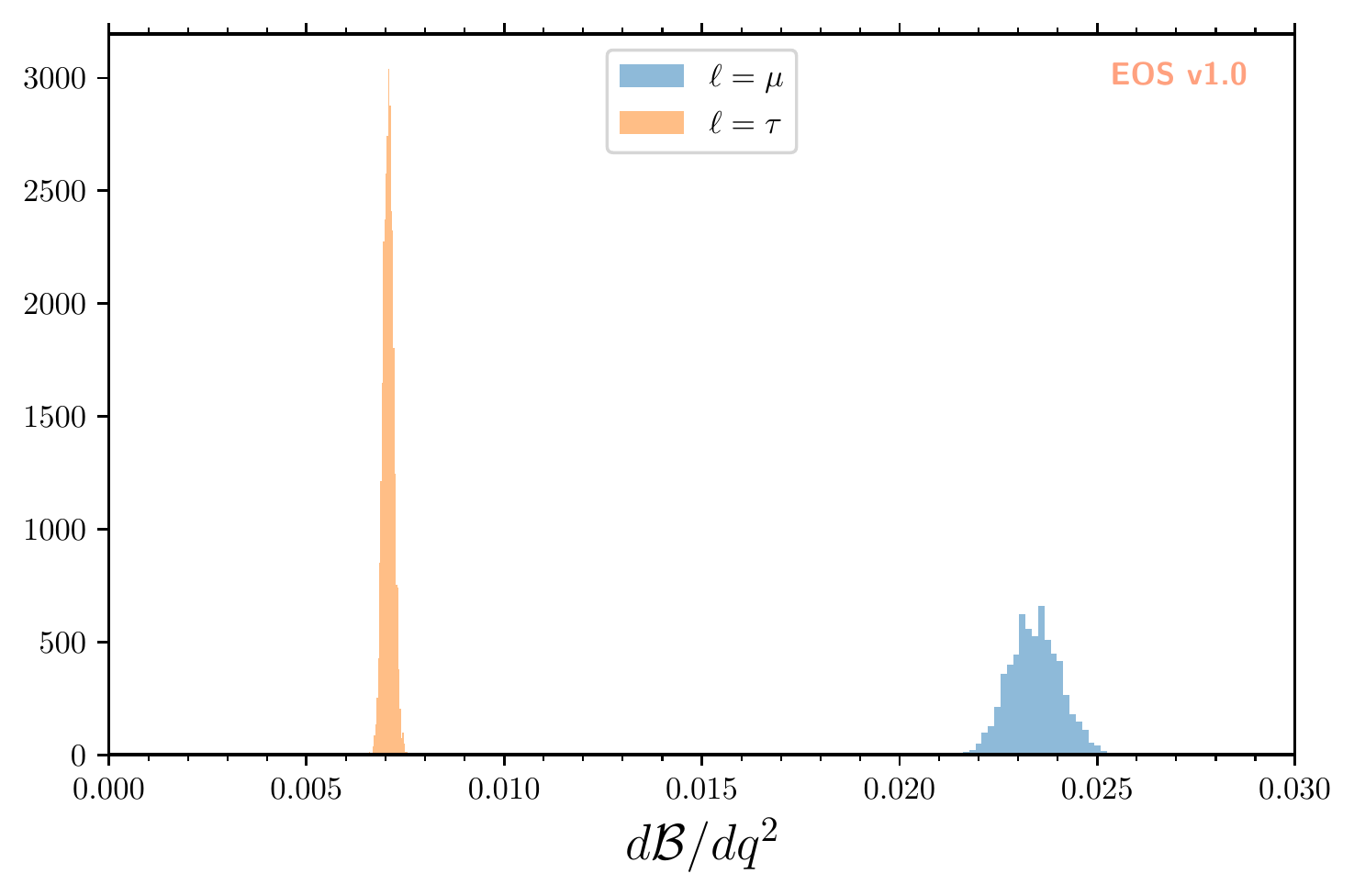}
    \includegraphics[width=0.48\linewidth]{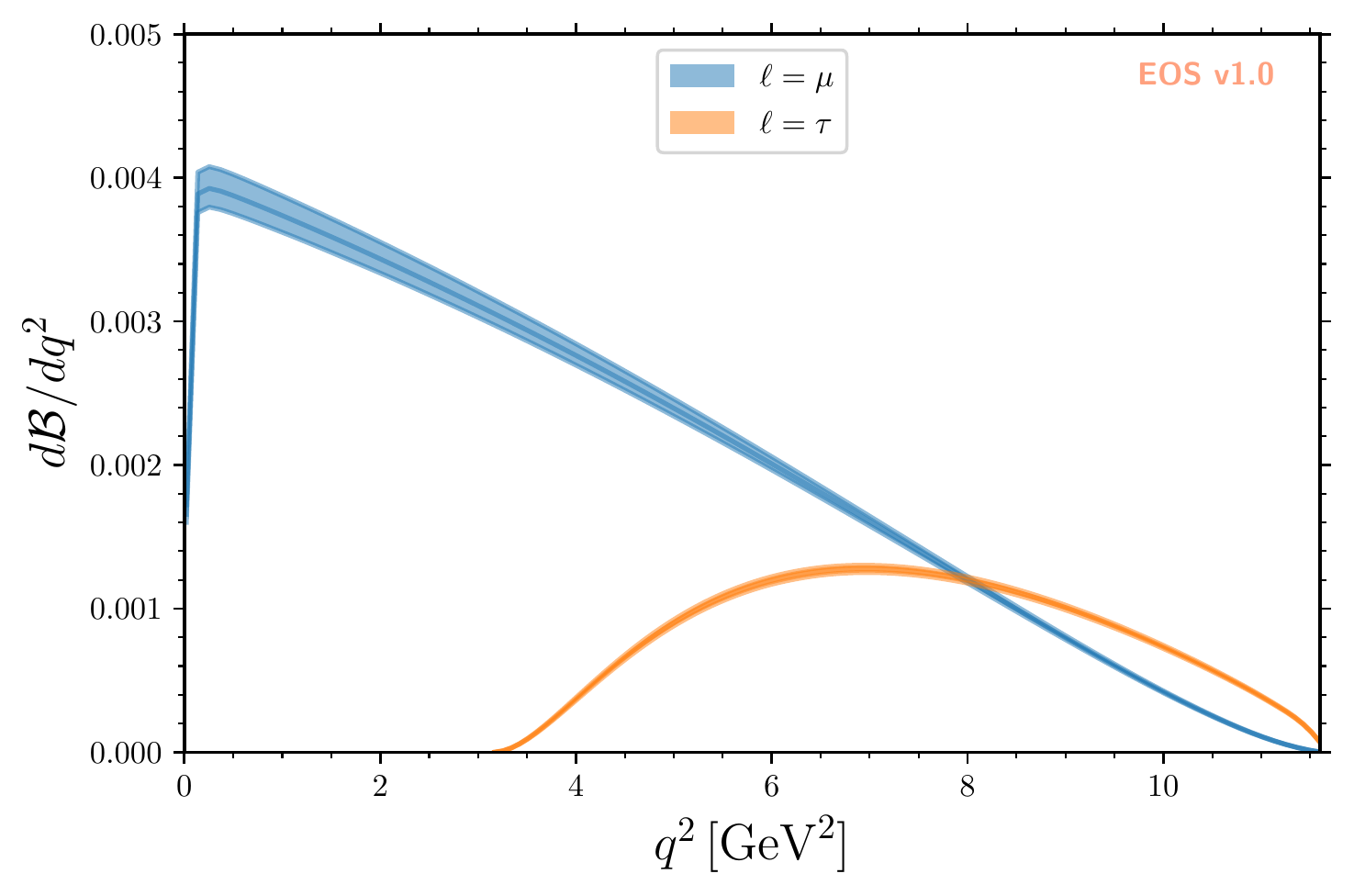}
    \caption{%
        Plot of the branching ratios of $\bar{B}\to D\lbrace \mu^-,\tau^-\rbrace\bar\nu$.
        Left: prior-predictive samples for the integrated branching ratios obtained from the code in \reflst{usage:prior-samples-int}.
        Right: differential branching ratios as functions of $q^2$. The central curves are obtained from \reflst{usage:BtoDlnu:BR}.
        The uncertainty bands are obtained from the samples obtained in \reflst{usage:prior-samples-diff}
        using the plotting code in \reflst{usage:plot-prior-prediction-diff}.
    }
    \label{out:usage:prior-prediction}
\end{joutput}

%
%
%--------+---------+---------+---------+---------+---------+---------+---------+
\subsection{Parameter Inference}
\label{sec:usage:inference}

[\textit{The example developed in this section can be ran interactively from the example notebook for parameter inference available
from ref.~\cite{EOS:repo}, file \href{https://github.com/eos/eos/tree/v1.0/examples/inference.ipynb}{examples/inference.ipynb}}]\\

\EOS infers parameters from a database of experimental or theoretical constraints in combination with its built-in observables.
This section illustrates how to construct an \class{eos.Analysis} object that represents the statistical analysis
and to infer the best-fit point and uncertainties of a list of parameters through optimization and Monte Carlo methods.
We pick up the example introduced in \refsec{basics:analysis} to illustrate the above-mentioned features of \EOS.
In particular, we use the two experimental constraints \code{B^0->D^+e^-nu::BRs@Belle:2015A} and \code{B^0->D^+mu^-nu::BRs@Belle:2015A}, to infer the value of the \CKM matrix element $|V_{cb}|$.

%
%
%--------+---------+---------+---------+---------+---------+---------+---------+
\subsubsection{Defining the Statistical Analysis} \label{sec:usage:analysis}

To define our statistical analysis for the inference of $|V_{cb}|$ from measurements of the $\bar{B}\to D\ell^-\bar\nu$
branching ratios, some decisions are needed.
First, we must decide how to parametrize the hadronic form factors that describe semileptonic $\bar{B}\to D$ transitions.
For what follows we will use the parametrization of Ref. \cite{Straub:2015ica}, referred to as \code{[BSZ:2015A]}.
Next, we must decide the theory input for the form factors.
For this, we will combine the correlated lattice QCD results published by the Fermilab/MILC and HPQCD collaborations in 2015 \cite{Na:2015kha,MILC:2015uhg}.

The corresponding \object{eos.Analysis} object is shown in \reflst{basics:analysis:definition}; it has been used previously
as an example in \refsec{basics:analysis}.
The global options ensure that our choice of form factor parametrization is used throughout, and that for \CKM matrix elements the \code{CKM} model is used.
The latter provides parametric access to the $V_{cb}$ matrix element through two objects of type \object{eos.Parameter}:
the absolute value \code{CKM::abs(V_cb)} and the complex phase \code{CKM::arg(V_cb)}.
The latter is not accessible from $b\to c\ell\bar\nu$.

We also set the starting value of \code{CKM::abs(V_cb)} to a sensible value of $42 \ng*{\cdot} 10^{-3}$ \textit{via}
\begin{lstlisting}[language=iPython]
analysis.parameters['CKM::abs(V_cb)'].set(42.0e-3)
\end{lstlisting}

\begin{joutput}[t]
    \resizebox{\textwidth}{!}{%
    \begin{tabular}{ll}
    \toprule
        parameter & value \\
    \midrule
        $|V_{cb}|$ & 0.0422 \\
        % use \texttt{} to avoid trouble...
        \texttt{B->D::alpha\^{}f+\_0@BSZ2015} & 0.6671 \\
        \texttt{B->D::alpha\^{}f+\_1@BSZ2015} & -2.5314 \\
        \texttt{B->D::alpha\^{}f+\_2@BSZ2015} & 4.8813 \\
        \texttt{B->D::alpha\^{}f0\_1@BSZ2015} & 0.2660 \\
        \texttt{B->D::alpha\^{}f0\_2@BSZ2015} & -0.8410 \\
    \bottomrule
    \end{tabular}
    \hspace*{1em}
    \begin{tabular}{lll}
    \toprule
         constraint & $\chi^2$ & d.o.f. \\
    \midrule
         % use \texttt{} to avoid trouble...
         \texttt{B->D::f\_++f\_0@HPQCD:2015A} & 3.4847 & 7 \\
         \texttt{B->D::f\_++f\_0@FNAL+MILC:2015B} & 3.1016 & 5 \\
         \texttt{B\^{}0->D\^{}+e\^{}-nu::BRs@Belle:2015A} & 11.8206 & 10 \\
         \texttt{B\^{}0->D\^{}+mu\^{}-nu::BRs@Belle:2015A} & 5.2242 & 10 \\
    \bottomrule
    \end{tabular}
    \hspace*{1em}
    \begin{tabular}{ll}
    \toprule
         total $\chi^2$ & 23.6310 \\
         total degrees of freedom & 26 \\
         p-value & 59.7053\% \\
    \bottomrule
    \end{tabular}
    }
    \caption{%
        Display of the best-fit point and goodness-of-fit summary obtained from optimizing the
        the $\bar{B}\to D\ell^-\bar\nu$ analysis shown in \reflst{basics:analysis:definition}.
    }
    \label{out:inference:bfpgof}
\end{joutput}

To maximize the (logarithm of the) posterior density we can call the \method{eos.Analysis}{optimize} method, as shown in \reflst{inference:bfpgof}.
In a \Jupyter notebook, it is useful to display the return value of this method, which illustrates the best-fit point.
Further useful information is contained in the goodness-of-fit summary.
The latter lists each constraint, its degrees of freedom, and its $\chi^2$ value (if applicable\footnote{%
    Note that \EOS supports likelihood functions
    that do not have a $\chi^2$ test statistic or any test statistic at all.
}), alongside the $p$-value for the entire likelihood.
\begin{lstlisting}[%
    language=iPython,%
    caption={%
        Optimize the posterior density and provide the best-fit point and goodness-of-fit summary.
        The output is shown in \refout{inference:bfpgof}.
        \label{lst:inference:bfpgof}
    }%
]
bfp = analysis.optimize()
display(bfp)
display(analysis.goodness_of_fit())
\end{lstlisting}
Instead of setting individual parameters to sensible values as we did for \code{CKM::abs(V_cb)} earlier,
a starting point can alternatively be provided to \method{eos.Analysis}{optimize} using the \code{start_point} keyword argument.

The maximization of the posterior by means of \method{eos.Analysis}{optimize} uses \SciPy's \pyclass{optimize}
module~\cite{scipy}. The default optimization algorithm is the Sequential Least SQuares Programming (SLSQP).
Other algorithms can be selected and configured through keyword arguments that \method{eos.Analysis}{optimize}
forwards to \pyclass{scipy.optimize}.\\

To interface with optimizers other than available within \SciPy, \EOS provides the \method{eos.Analysis}{log\_pdf} method.
As its first argument, it expects the list of the parameter values. The parameters' ordering must correspond to
the ordering of \object{analysis.varied_parameters}, and each parameter's values must be rescaled to the interval $[-1, +1]$,
where the boundaries correspond to the minimal/maximal value in the prior specification.

\subsubsection{Importance Sampling of the Posterior}

To sample from the posterior, \EOS provides the \method{eos.Analysis}{sample} method.
Optionally, this can also produce posterior-predictive samples for a list of observables.
We can use these samples to illustrate the results of our fit in relation to the experimental constraints.\\

For this example, we produce such posterior-predictive samples for the differential $\bar{B}\to D^+\mu^-\bar\nu$ branching
ratio in the 40 points of the kinematic variable $q^2$ used in the previous examples (redifined in the following
listing for completeness).

\begin{lstlisting}[%
    language=iPython,%
    caption={%
        Produce posterior-predictive samples for the differential $\bar{B}\to D^+\mu^-\bar\nu$ branching ratio. \label{lst:inference:posterior_predictive_sample_distribution}
    }%
]
mu_q2values = numpy.unique(numpy.concatenate((numpy.linspace(0.02,  1.00, 20),
                                              numpy.linspace(1.00, 11.60, 20))))
mu_obs      = [eos.Observable.make('B->Dlnu::dBR/dq2', analysis.parameters, eos.Kinematics(q2=q2),
                                   eos.Options({'form-factors': 'BSZ2015', 'l': 'mu', 'q': 'd'})) for q2 in mu_q2values]
parameter_samples, log_weights, mu_samples = analysis.sample(N=20000, stride=5, pre_N=1000, preruns=5, start_point=bfp.point,
                observables=mu_obs)
\end{lstlisting}

In the above we start sampling at the best-fit point as obtained earlier through optimization, which is optional.
We carry out 5 burn-in runs/preruns of 1000 samples each.
The samples obtained in each of these preruns are used to adapt the Markov chain but are then discarded.
The main run produces a total of \code{N * stride = 100000} random Markov Chain samples.
The latter are thinned down by a factor of \code{stride = 5} to obtain \code{N = 20000} samples, which are stored in \code{parameter_samples}.
The thinning reduces the autocorrelation of the samples.
The values of the log(posterior) are stored in \code{log_posterior}.
The posterior-predictive samples for the observables are stored in \code{e_samples}, and are only returned if the observables keyword argument is provided.\\

We can now illustrate the posterior samples either as a histogram or as a \KDE using the built-in plotting functions,
see \refout{inference:posterior-sample-hist+kde} and \reflst{plot-ex:inference:posterior-sample-hist}.
Contours at given levels of posterior probability, as shown in \refout{inference:posterior-sample-hist+kde},
can be obtained for any pair of parameters using \reflst{plot-ex:inference:posterior-sample-kde}.\\

\begin{joutput}[t]
    \centering
    \includegraphics[width=0.48\linewidth]{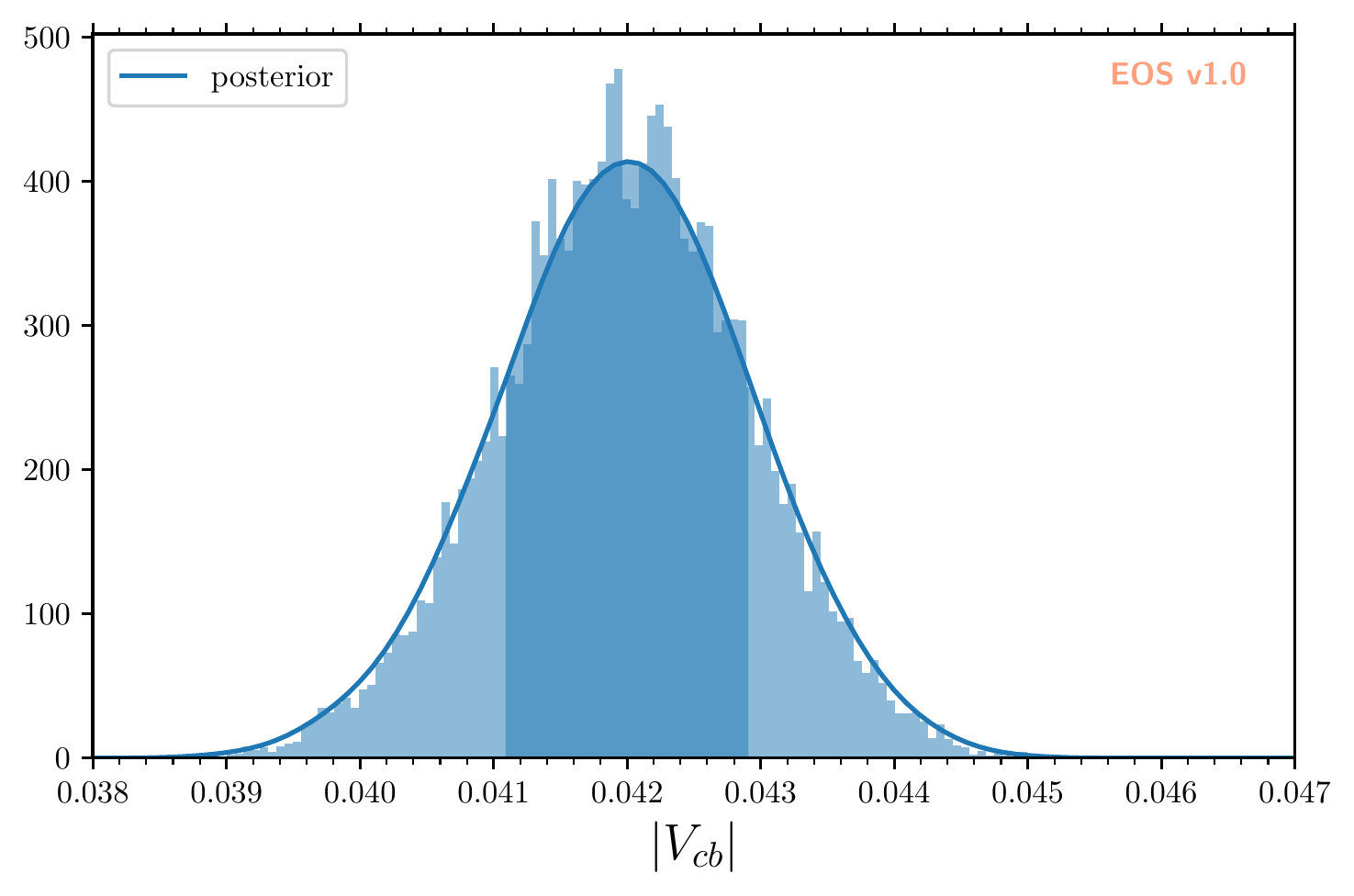}
    \includegraphics[width=0.48\linewidth]{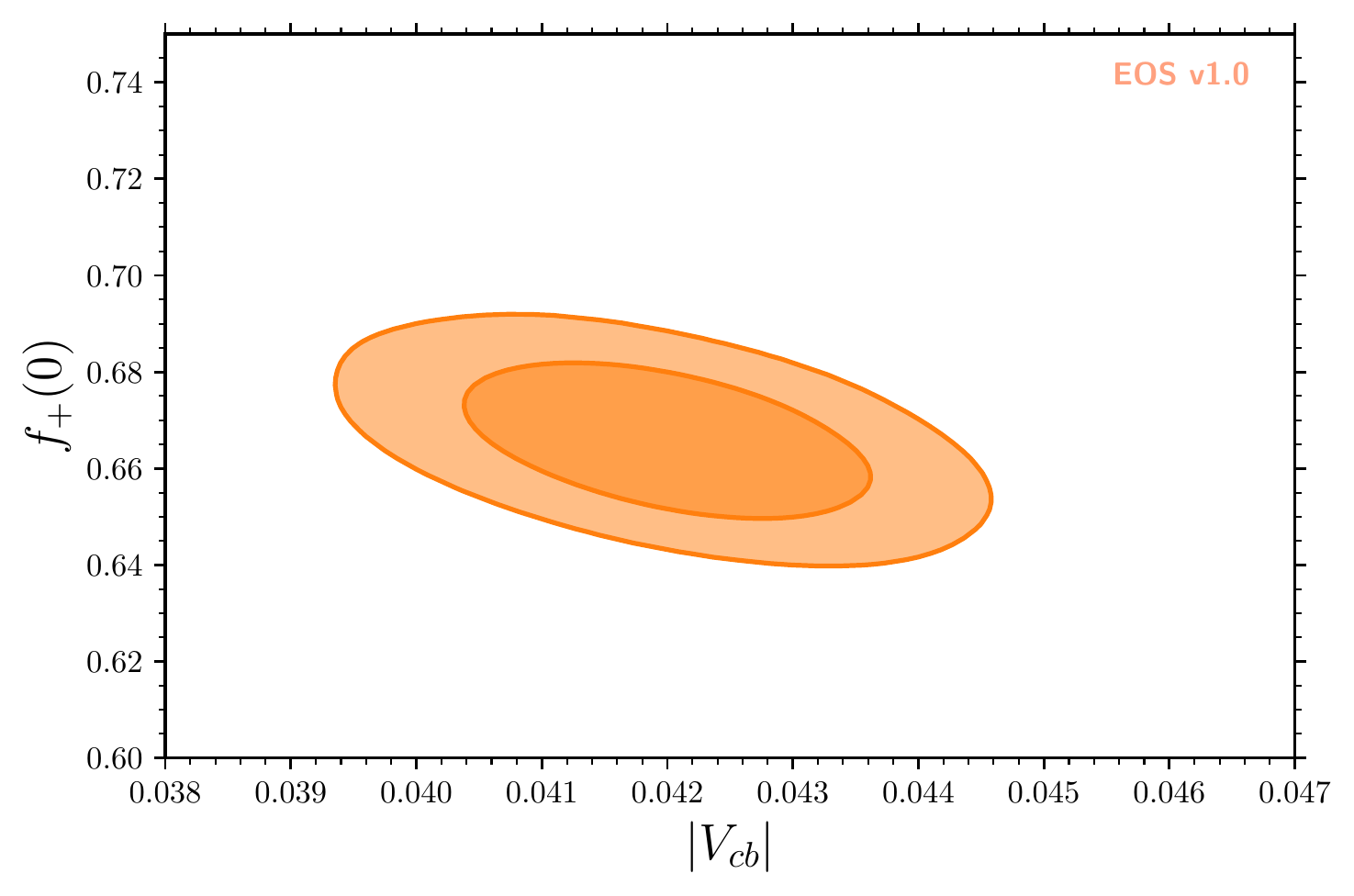}
    \caption{%
        Distribution of samples (left) of the 1D-marginal posterior of $|V_{cb}|$ as a regular histogram
        and as a kernel density estimate (blue line); and (right) of the 2D-marginal joint posterior
        of $|V_{cb}|$ and $f^{\bar{B}\to D}_+(0)$ as contours at $68\%$ and $95\%$ probability (orange lines and filled areas).
        The plots are produced by \reflst{plot-ex:inference:posterior-sample-hist}
        and \reflst{plot-ex:inference:posterior-sample-kde}, respectively.
    }
    \label{out:inference:posterior-sample-hist+kde}
\end{joutput}

Sampling with the Metropolis-Hastings algorithm is known to work well for unimodal densities.
However, in cases of multimodal densities or blind directions, problems regularly arise.
\EOS provides the means to follow the approach of ref.~\cite{2013arXiv1304.7808B}, which
proposes to use (potentially unadapted) Markov chains to explore the parameter space to
initialize a Gaussian mixture density. The latter is then adapted using the Population
Monte Carlo algorithm~\cite{2010MNRAS.405.2381K}, for which \EOS uses the \pypmc package~\cite{pypmc}.
Within \EOS, we use schematically the following approach:
\begin{lstlisting}[
    language=iPython,%
    caption={%
        Create a mixture density from a number of Markov chains, and adapt it to the posterior
        through a call to \code{eos.Analysis.sample_pmc}\index{eos.Analysis!sample\_pmc}.
        \label{lst:inference:sample_pmc}
    }%
]
from pypmc.mix_adapt.r_value import make_r_gaussmix
chains = []
for i in range(10):
  # run Markov Chains for your problem
  chain, _ = analysis.sample(...) # use relevant settings for your analysis in the '...'
  chains.append(chain)

# please consult the pypmc documentation for details on the call below
proposal_density = make_r_gaussmix(chains, K_g=3, critical_r=1.1)

# adapt the proposal to the posterior and obtain high-quality samples
analysis.sample_pmc(proposal_density, ...) # use relevant settings for your analysis in the '...'
\end{lstlisting}

We can visualize the posterior-predictive samples using:
\begin{lstlisting}[%
    language=iPython,%
    caption={%
        Plot posterior-predictive importance samples for the differential $\bar{B}\to D^+\mu^-\bar\nu$ branching ratio vs. $q^2$.
        The result is shown in \refout{inference:posterior-prediction-diff}.
        \label{lst:inference:posterior_samples_uncertainties}\index{eos.Plotter!plot}
    }%
]
plot_args = {
    'plot': {
        'x': { 'label': r'$q^2$', 'unit': r'$\textnormal{GeV}^2$', 'range': [0.0, 11.63] },
        'y': { 'label': r'$d\mathcal{B}/dq^2$',                    'range': [0.0,  5e-3] },
        'legend': { 'location': 'lower left' }
    },
    'contents': [
        {
            'label': r'$\ell=\mu$',
            'type': 'uncertainty',
            'range': [0.02, 11.60],
            'data': { 'samples': mu_samples, 'xvalues': mu_q2values }
        },
        {
            'label': r'Belle 2015 $\ell=e,\, q=d$',
            'type': 'constraint',
            'color': 'C0',
            'constraints': 'B^0->D^+e^-nu::BRs@Belle:2015A',
            'observable': 'B->Dlnu::BR',
            'variable': 'q2',
            'rescale-by-width': True
        },
        {
            'label': r'Belle 2015 $\ell=\mu,\,q=d$',
            'type': 'constraint',
            'color': 'C1',
            'constraints': 'B^0->D^+mu^-nu::BRs@Belle:2015A',
            'observable': 'B->Dlnu::BR',
            'variable': 'q2',
            'rescale-by-width': True
        },
    ]
}
eos.plot.Plotter(plot_args).plot()
\end{lstlisting}
Note that the use of \code{'rescale-by-width': True} converts the database's existing entry
for the \emph{bin-integrated} branching ratio into the \emph{bin-averaged} branching ratio.
Only that latter can be meaningfully compared with the differential branching ratio's curve.

\FloatBarrier
\begin{joutput}[t]
    \centering
    \includegraphics[width=0.48\linewidth]{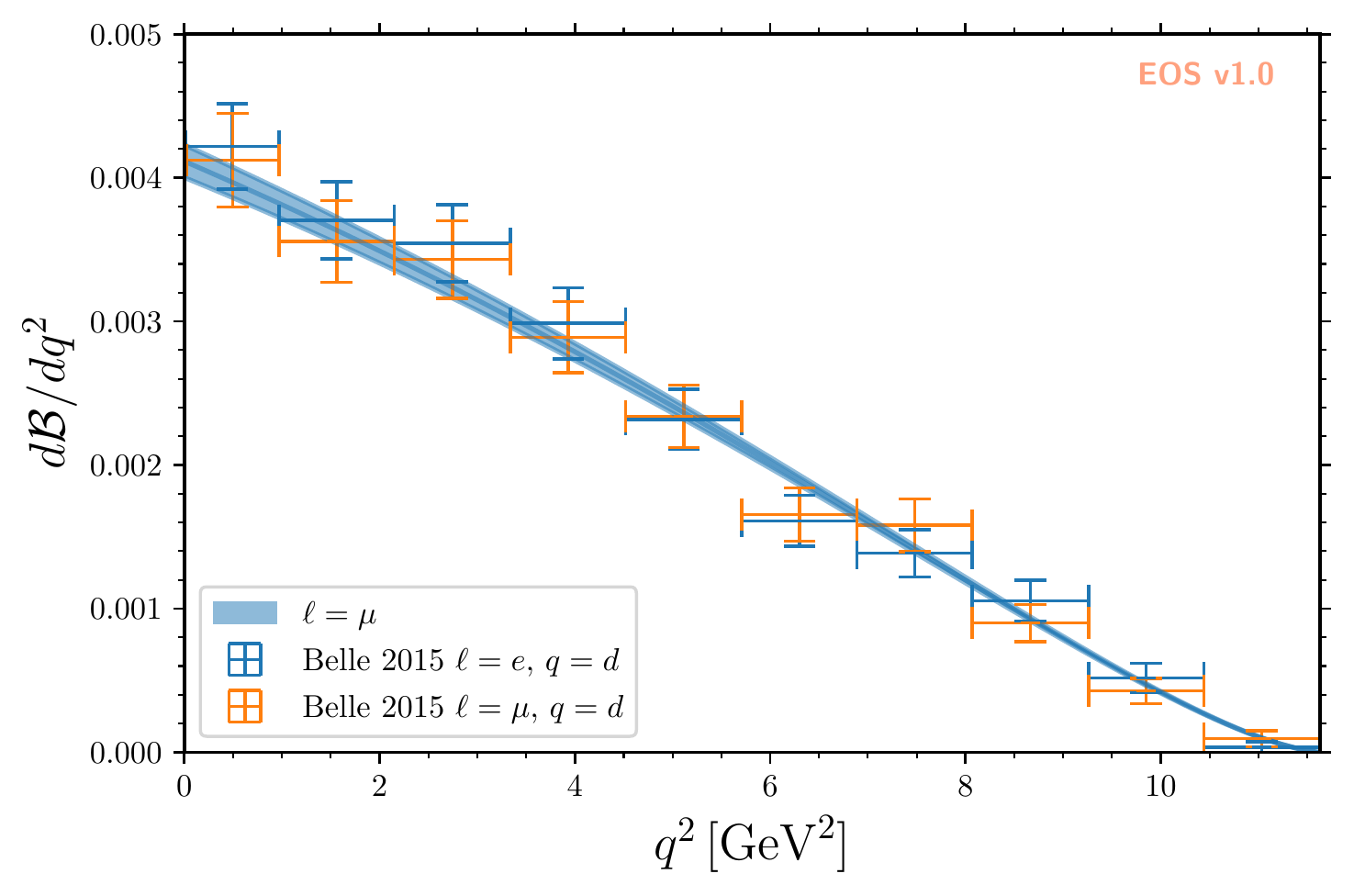}
    \caption{
        Plot of the posterior-predictive importance samples for the differential $\bar{B}\to D^+\mu^-\bar\nu$ branching ratio vs. $q^2$,
        juxtaposed with bin-averaged measurements of the $\bar{B}\to D^+\lbrace e^-,\mu^-\rbrace\bar\nu$ branching ratio
        by the Belle experiment.
    }
    \label{out:inference:posterior-prediction-diff}
\end{joutput}

%
%
%--------+---------+---------+---------+---------+---------+---------+---------+
\subsection{Event Simulation}
\label{sec:usage:simulation}

[\textit{The example developed in this section can be run interactively from the example notebook for event simulation available
from ref.~\cite{EOS:repo}, file \href{https://github.com/eos/eos/tree/v1.0/examples/simulation.ipynb}{examples/simulation.ipynb}}]\\

\EOS contains built-in probability density functions (PDFs) from which pseudo events can be simulated using Markov chain Monte Carlo techniques.

\subsubsection{Constructing a 1D PDF and Simulating Pseudo Events}

The simulation of events is performed using the \method{eos.SignalPDF}{sample\_mcmc} method.
For example, the construction of the one-dimensional PDF describing the $B\to D\ell\nu_\ell$ decay distribution in the variable $q^2$ and for $\ell=\mu$ leptons requires:
\begin{itemize}
    \item the \code{q2} kinematic variable that can be set to an arbitrary starting value.
    \item the boundaries, \code{q2_min} and \code{q2_max}, for the phase space from which we want to sample. If needed, the phase space can be shrunk to a volume smaller than physically allowed; the normalization of the PDF will automatically adapt.
\end{itemize}

For $B\to D\ell\nu_\ell$, the Markov chains can self adapt to the PDF in 3 preruns with 1000 pseudo events/samples each.

The simulation of \code{stride*N=250000} pseudo events/samples from the PDF, which are thinned down to \code{N=50000}, is performed with the following code:
\begin{lstlisting}[%
    language=iPython,%
    caption={%
        Produce importance samples of the one-dimensional \code{SignalPDF}
        for the $\bar{B}\to D\ell^-\bar\nu$ differential branching ratio.
        The samples are compared to the analytic expression in the left figure of \refout{simulation:plot+histogram},
        which is produced from \reflst{plot-ex:simulation:plot+histogram-1D}.
        \label{lst:simulation:sample-1D}
    }%
]
rng = numpy.random.mtrand.RandomState(123456) # Defines a seeded random number generator
mu_kinematics          = eos.Kinematics({'q2': 2.0, 'q2_min': 0.02, 'q2_max': 11.6})
mu_options             = eos.Options({'l': 'mu'})
mu_pdf                 = eos.SignalPDF.make('B->Dlnu::dGamma/dq2', eos.Parameters(), mu_kinematics, mu_options)
mu_samples, mu_weights = mu_pdf.sample_mcmc(N=50000, stride=5, pre_N=1000, preruns=3, rng=rng)
\end{lstlisting}
Samples for other lepton flavors, e.g., $\ell=\tau$, require only a change of the \class{eos.Options} object to use \code{'l': 'tau'} instead
and adjustment of the phase space.
Similar to observables, \class{eos.SignalPDF} objects can be plotted as a function of a single kinematic variable, while keeping all other kinematic variables fixed.
The fixed kinematic variables are provided as a \pyclass{dict} via the \code{kinematics} key.
We show two such plots in combination with histograms of the \PDF samples in \refout{simulation:plot+histogram} (left).
The output shows excellent agreement between the simulations and the respective analytic expressions for the 1D \acp{PDF}.

\subsubsection{Constructing a 4D PDF and Simulating Pseudo Events}

Samples can also be drawn for \acp{PDF} with more than one kinematic variable.
As an example, we use the full four-dimensional \PDF for $\bar{B}\to D^*\ell\bar{\nu}$ decays.
Declaration and initialization of all four kinematic variables
(\code{q2}, \code{cos(theta_l)}, \code{cos(theta_d)}, and \code{phi}) is similar to the 1D case.
\begin{lstlisting}[language=iPython,
    caption={%
        Produce importance samples of the four-dimensional \code{SignalPDF}
        for the $\bar{B}\to D^*(\to D\pi)\ell^-\bar\nu$ differential branching ratio.
        \label{lst:simulation:sample-4D}
    }%
]
dstarlnu_kinematics = eos.Kinematics({
    'q2':            2.0,  'q2_min':            0.02,     'q2_max':           10.5,
    'cos(theta_l)':  0.0,  'cos(theta_l)_min': -1.0,      'cos(theta_l)_max': +1.0,
    'cos(theta_d)':  0.0,  'cos(theta_d)_min': -1.0,      'cos(theta_d)_max': +1.0,
    'phi':           0.3,  'phi_min':           0.0,      'phi_max':           2.0 * numpy.pi
})
\end{lstlisting}
We then produce the samples in a similar way as for the 1D \PDF:
\begin{lstlisting}[language=iPython]
rng = numpy.random.mtrand.RandomState(74205) # Defines a seeded random number generator
dstarlnu_pdf        = eos.SignalPDF.make('B->D^*lnu::d^4Gamma', eos.Parameters(), dstarlnu_kinematics, eos.Options())
dstarlnu_samples, _ = dstarlnu_pdf.sample_mcmc(N=1e6, stride=5, pre_N=1000, preruns=3, rng=rng)
\end{lstlisting}
The samples of the individual kinematic variables can be accessed as the columns of the \code{dstarlnu_samples} object.
We can now show correlations of the kinematic variables by plotting 2D histograms, e.g. $q^2$ vs $\cos\theta_\ell$:
\begin{lstlisting}[%
    language=iPython,%
    caption={%
        Plot a 2D histogram for samples of the $\bar{B}\to D^*(\to D\pi)\mu^-\bar\nu$ PDF
        in the variables $q^2$ and $\cos(\theta_\ell)$.
        The samples are obtained from \reflst{simulation:sample-4D}, and the output is shown in the right plot of \refout{simulation:plot+histogram}.
        \label{lst:simulation:histogram-2D}\index{eos.Plotter!plot}
    }%
]
plot_args = {
    'plot': {
        'x': { 'label':r'$q^2$', 'unit': r'$\textnormal{GeV}^2$', 'range': [0.0, 10.50]},
        'y': { 'label':r'$cos(\theta_\ell)$',                     'range': [-1.0, +1.0]}
    },
    'contents': [
        {
            'label': r'samples ($\ell=\mu$)',
            'type': 'histogram2D',
            'data':{
                'samples': dstarlnu_samples[:, (0,1)]
            },
            'bins': 40
        },
    ]
}
eos.plot.Plotter(plot_args).plot()
\end{lstlisting}

\begin{joutput}[t]
    \centering
    \includegraphics[width=0.48\linewidth]{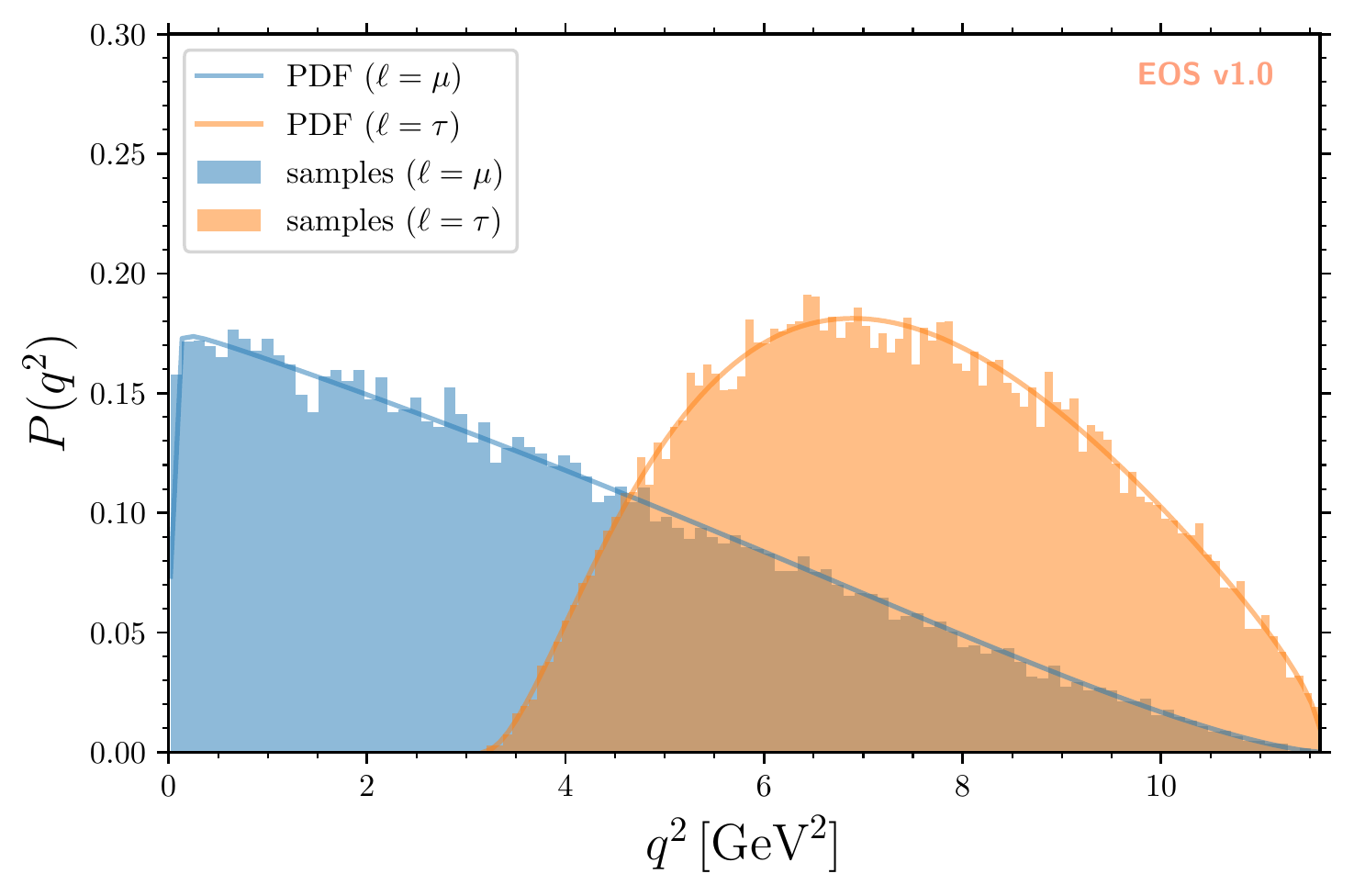}
    \includegraphics[width=0.48\linewidth]{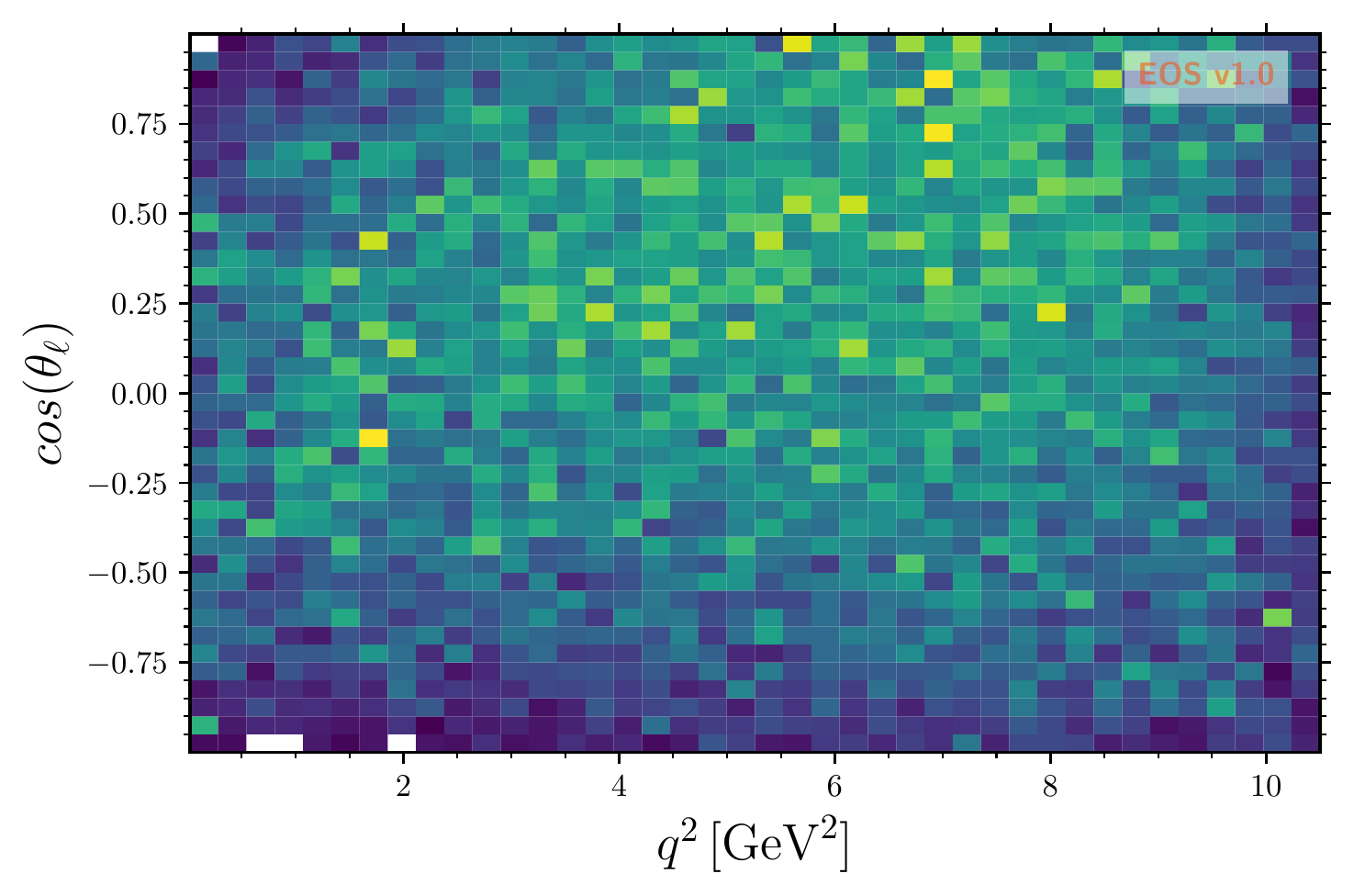}
    \caption{%
        Left: Distribution of $B\to D\ell\nu_\ell$ events for $\ell=\mu, \tau$, as implemented in \EOS (solid lines)
        and as obtained from Markov Chain Monte Carlo importance sampling (histograms).
        The samples are produced from \reflst{simulation:sample-1D}, and the plot is produced
        by \reflst{plot-ex:simulation:plot+histogram-1D}.
        Right: 2D histogram of the $\bar{B}\to D^*(\to D\pi)\mu^-\bar\nu$ PDF in the variables $q^2$ and $\cos(\theta_\ell)$.
        This output is produced by the code shown in \reflst{simulation:histogram-2D}.
    }
    \label{out:simulation:plot+histogram}
\end{joutput}

%
%
%--------+---------+---------+---------+---------+---------+---------+---------+
%
%
%--------+---------+---------+---------+---------+---------+---------+---------+
\section{Conclusion \& Outlook}\label{sec:summary}

We have presented the \EOS software in version \EOSversion and explained its three main use cases at the hand
of concrete examples in the field of flavor physics phenomenology.
Beyond these examples, \EOS has been used extensively for numerical evaluations, statistical analyses
and plots in a number of peer-reviewed publications. We plan to extend \EOS with further processes
and observables, while keeping the \Python interface unchanged.\\

To keep this document concise, some advanced aspects of \EOS have not been discussed.
These aspects, documented in the online documentation~\cite{EOS:doc}, include
\begin{itemize}
    \item the possibility to combine existing observables in arithmetic expressions at run time;
    \item the command line interface intended to use as part of massively parallelized batch jobs in grid or cluster environments; and
    \item the addition of \cpp code for new observables and processes.
\end{itemize}

Despite ongoing unit testing and development of the software, we are conscious that \EOS is
neither free of bugs nor providing all the features the user could possibly need. We therefore encourage
the users to report any and all bugs found and to request additional features.
We ask that any such reports or requests are communicated as issues within the \EOS Github
repository~\cite{EOS:repo}.
We are very happy to discuss the addition of further observables and processes with interested parties
from the phenomenological and experimental communities.

%
%
%--------+---------+---------+---------+---------+---------+---------+---------+
\section*{Acknowledgments}

DvD is grateful to Gudrun Hiller, Thomas Mannel, Gino Isidori, and Nico Serra,
whose support made the development of \EOS possible in the first place.
We thank all \EOS contributors who are not authors of this paper,
including Bastian M\"uller, Romy O'Connor, Stefanie Reichert, Martin Ritter, Eduardo Romero,
Ismo Toijala, and Christian Wacker.\\
% DvD (+EE, NG, SK, MR)
The work of DvD, EE, NG, SK, and MR and the development of EOS is supported by the German Research Foundation (DFG)
within the Emmy Noether Programme under grant DY 130/1-1 and by the National Natural Science Foundation of China (NSFC)
and the DFG through the funds provided to the Sino-German Collaborative Research Center TRR110
``Symmetries and the Emergence of Structure in QCD''
(NSFC Grant No. 12070131001, DFG Project-ID 196253076 -- TRR 110).
% FB: doesn't need acknowledgments anymore
% TB
The work of TB is supported by the Royal Society (United Kingdom).
% CB
The work of CB was supported by the DFG under grant BO-4535/1-1.
% MB, MJ
The work of MB and MJ is supported by the Italian Ministry of Research (MIUR) under
grant PRIN 20172LNEEZ.
% NG
The work of NG is also supported by the DFG under grant 396021762 -- TRR 257 ``Particle
Physics Phenomenology after the Higgs Discovery''.
% EG, RSC
The work of RSC and EG was supported by the Swiss National Science Foundation (SNSF) under contracts 159948, 172637 and 174182.
% PL
The work of PL is supported by the Cluster of Excellence ``ORIGINS'' funded by
the DFG under Germany's Excellence Strategy -- EXC-2094 -- 390783311.
% JV
The work of JV is supported by funding from the Spanish MICINN through the ``Ram\'on y Cajal'' program RYC-2017-21870,
the ``Unit of Excellence Mar\'ia de Maeztu 2020-2023” award to the Institute of Cosmos Sciences (CEX2019-000918-M) and from PID2019-105614GB-C21 and 2017-SGR-929 grants.
% KKV: doesn't need acknowledgments

%
%
%--------+---------+---------+---------+---------+---------+---------+---------+
\newpage
%--------+---------+---------+---------+---------+---------+---------+---------+
%
%
%
%--------+---------+---------+---------+---------+---------+---------+---------+
\appendix

\renewcommand{\thesection}{\Alph{section}}
\renewcommand{\thelstlisting}{\Alph{section}.\arabic{lstlisting}}
\numberwithin{lstlisting}{section}
\renewcommand{\sectcounterend}{}

%
%
%--------+---------+---------+---------+---------+---------+---------+---------+
\section{Details on the Models}
\label{app:models}

\subsection{The CKM Model and Parameters}
\label{app:models:CKM}

When evaluating low-energy observables within the \SM, the \CKM matrix elements
are evaluated using the Wolfenstein parametrization~\cite{Wolfenstein:1983yz} expanded to
order $\lambda^8$~\cite{Charles:2004jd}. The Wolfenstein parameters $\lambda$
and $A$ are used without modifications. The $\rho$ and $\eta$ parameters are traded
for $\bar{\rho}$ and $\bar{\eta}$~\cite{Charles:2004jd}, the coordinates of the
apex of the standard unitarity triangle. The two parameters are defined to all orders
in $\lambda$ as~\cite{Charles:2004jd}
\begin{equation}
\begin{aligned}
    \bar\rho & = -\Re \frac{V_{ud}^{\phantom{*}}\,V_{ub}^*}{V_{cd}^{\phantom{*}}\,V_{cb}^*}\,, &
    \bar\eta & = -\Im \frac{V_{ud}^{\phantom{*}}\,V_{ub}^*}{V_{cd}^{\phantom{*}}\,V_{cb}^*}\,.
\end{aligned}
\end{equation}
These four parameter can be accessed \textit{via} \code{CKM::lambda}, \code{CKM::A},
\code{CKM::rhobar}, and \code{CKM::etabar}, respectively.\\

A frequent physics use case involves inferring the absolute value or complex argument
of a \CKM matrix element from data. Choosing the \CKM model using \code{'model': 'CKM'}
ensures that each complex-valued \CKM matrix element is parametrized in terms of
its absolute value and complex argument. For example, the parametrization
of the \CKM matrix element $V_{ub}$ involves the parameters
\code{CKM::abs(V_ub)} and \code{CKM::arg(V_ub)}. The names for the remaining \CKM
parameter follow the same naming scheme.

\subsection{The WET Model, Operator Bases, and Parameters}
\label{app:models:WET}

The observables for low-energy processes below the electroweak scale rely on
a description in the \WET, both within the 
\SM~\cite{Buchalla:1995vs,Buras:1998raa,Buras:2011we} and in \BSM scenarios
\cite{Aebischer:2017gaw,Jenkins:2017jig}. Observables can be evaluated
within the \WET by setting the \code{model} option to \code{WET}.
Within this model, \WET Wilson coefficients are parametrized
by individual \EOS parameters, and \CKM matrix elements are treated as
in the \code{CKM} model; see \refapp{models:CKM}.

Within \EOS, the \WET is parametrized as
\begin{equation}
    \mathcal{L}^{\textrm{WET}} = \sum_{\mathcal{S}} \mathcal{L}^{\mathcal{S}}\,,
\end{equation}
where $\mathcal{S}$ denotes a \emph{sector} of the \WET, \emph{i.e.}, a set of
operators with definite quantum numbers under global symmetries preserved by
the renormalization group evolution~\cite{Aebischer:2017ugx}.
For each sector, \EOS follows the \WCxf convention~\cite{Aebischer:2017ugx}:
\begin{equation}
    \mathcal{L}^{\mathcal{S}} \equiv 
        \sum_{\mathcal{O}_i^{\mathcal{S}} \neq \mathcal{O}_i^{\mathcal{S},\dagger}}
        \left[\mathcal{C}^\mathcal{S}_i \, \mathcal{O}^{\mathcal{S}}_i + \text{h.c.}\right]
    +   \sum_{\mathcal{O}_i^{\mathcal{S}} = \mathcal{O}_i^{\mathcal{S},\dagger}}
        \mathcal{C}^\mathcal{S}_i \, \mathcal{O}^{\mathcal{S}}_i\,.
\end{equation}
Here $\mathcal{O}$ denotes the dimension-six operators, and $\mathcal{C}$ is a
dimensionless Wilson coefficient renormalized at an appropriate low-energy scale $\mu$.
This scale is accessible as an \class{eos.Parameter}. The prefix part of its
qualified name corresponds to the sector and the name part of its qualified name
is \texttt{mu}, i.e., for the sector \texttt{sbsb} this parameter is named
\texttt{sbsb::mu}.\\

As of version 1.0, \EOS supports the following sectors:
\begin{itemize}
    \item \texttt{sb}
    \item \texttt{sbee}, \texttt{sbmumu}, \texttt{sbtautau}
    \item \texttt{sbnunu}
    \item \texttt{sbsb}
    \item \texttt{cbenue}, \texttt{cbmunumu}, \texttt{cbtaunutau}
    \item \texttt{ubenue}, \texttt{ubmunumu}, \texttt{ubtaunutau}
\end{itemize}
Changes to the parameters representing the Wilson coefficients only affect observables
that are constructed with the \code{model} option set to \code{'WET'}. 
By convention the Wilson coefficients comprise both their \SM value and
potential \BSM shifts, i.e.:
\begin{align}
    \mathcal{C}_i(\mu) &
    = \mathcal{C}_i^\text{SM}(\mu) + \mathcal{C}_i^\text{BSM}(\mu)\,.
\end{align}
A complete list of the sectors and their operators supported by \EOS is part
of the \WCxf basis file \cite{wcxf:EOS-basis}.
The parameters describing the Wilson coefficients are listed as part of the
\EOS documentation~\cite[List of Parameters]{EOS:doc}. Using \WCxf and
the \wilson package~\cite{Aebischer:2018bkb}, constraints on the \WET can be readily interpreted
as constraints at a different scale in the \WET or as constraints on
Wilson coefficients in the Standard Model Effective Field Theory~\cite{Buchmuller:1985jz,Grzadkowski:2010es}.

The \SM values of the \WET Wilson Coefficients up to mass dimension six are
known to high precision in the \SM. By default, \EOS evaluates observables
with the \code{model} option set to \code{'SM'}.
This choice of option leads \EOS to compute the WET Wilson Coefficients
at the electroweak scale $\mu_0$ and evolve them to the appropriate
low-energy scale $\mu$~\footnote{%
    The choice of $\mu_0$ should, in general, not be changed,
    and for some sectors there is more than one high scale involved.
}.
\begin{itemize}
    \item For the sectors \code{sb}, \code{sbee}, \code{sbmumu}, and \code{sbtautau},
    the SM values are computed to NNLO in QCD~\cite{Adel:1993ah,Greub:1997hf,Bobeth:1999mk}.
    The RG evolution to the low-energy scale $\sim m_b$ crucially requires the resummation of radiative QCD and partially
    also QED corrections~\cite{Chetyrkin:1996vx,Bobeth:2003at,Gorbahn:2004my,Gorbahn:2005sa,Huber:2005ig}.
    
    \item For the sector \code{sbnunu}, the SM values are computed to NLO in QCD~\cite{Misiak:1999yg,Buchalla:1998ba}.

    \item For the sectors \code{cbenue} through \code{ubtaunutau}, the SM values are
    computed to next-to-leading order in QED~\cite{Sirlin:1981ie}.
    
    \item For the sector \code{sbsb}, the SM values are computed to NLO in
    QCD~\cite{Buras:1990fn} and NLO in EW~\cite{Gambino:1998rt}.
    The RG evolution to the low scale $\sim m_b$ crucially requires the resummation
    of radiative QCD corrections~\cite{Buras:2000if}.
\end{itemize}

%--------+---------+---------+---------+---------+---------+---------+---------+
%
%
%--------+---------+---------+---------+---------+---------+---------+---------+
\section{Collection of Examples}
\label{app:PlotExamples}

Here we collect a number of code examples that are used in the main text to produce
a variety of plots. They have been moved to this appendix to ease legibility of the main text.

\begin{lstlisting}[%
    language=iPython,%
    caption={%
        Histogram samples of the 1D-marginal posterior for $|V_{cb}|$ and plot their kernel density estimate.
        This code is used to produce \refout{inference:posterior-sample-hist+kde} (left).
        \label{lst:plot-ex:inference:posterior-sample-hist}
    }
]
plot_args = {
    'plot': {
        'x': { 'label': r'$|V_{cb}|$', 'range': [38e-3, 47e-3] },
        'legend': { 'location': 'upper left' }
    },
    'contents': [
        {
            'type': 'histogram',
            'data': { 'samples': parameter_samples[:, 0] }
        },
        {
            'type': 'kde', 'color': 'C0', 'label': 'posterior', 'bandwidth': 2,
            'range': [38e-3, 47e-3],
            'data': { 'samples': parameter_samples[:, 0] }
        }
    ]
}
eos.plot.Plotter(plot_args).plot()
\end{lstlisting}

\begin{lstlisting}[%
    language=iPython,%
    caption={%
        Plot contours of the joint 2D-marginal posterior of the parameters $|V_{cb}|$ and $f_+(0)$
        at $68\%$ and $95\%$ probability using a kernel density estimate.
        This code is used to produce \refout{inference:posterior-sample-hist+kde} (right).
        \label{lst:plot-ex:inference:posterior-sample-kde}
    }
]
plot_args = {
    'plot': {
        'x': { 'label': r'$|V_{cb}|$', 'range': [38e-3, 47e-3] },
        'y': { 'label': r'$f_+(0)$',   'range': [0.6, 0.75] },
    },
    'contents': [
        {
            'type': 'kde2D', 'color': 'C1', 'label': 'posterior',
            'levels': [68, 95], 'contours': ['lines', 'areas'], 'bandwidth': 3,
            'data': { 'samples': parameter_samples[:, (0,1)] }
        }
    ]
}
eos.plot.Plotter(plot_args).plot()
\end{lstlisting}

\begin{lstlisting}[%
    language=iPython,%
    caption={%
        Plot the analytical 1D PDFs and 1D-marginal histograms of pseudo events
        for the decays $\bar{B}\to D\lbrace \mu^-,\tau^-\rbrace \bar{\nu}$.
        The pseudo events for the semimuonic decay are obtained from \reflst{simulation:sample-1D}.
        The result is shown in the left plot of \refout{simulation:plot+histogram}.
        \label{lst:plot-ex:simulation:plot+histogram-1D}
    }
]
plot_args = {
    'plot': {
        'x': {'label': r'$q^2$', 'unit': r'$\textnormal{GeV}^2$', 'range': [0.0, 11.60]},
        'y': {'label': r'$P(q^2)$', 'range': [0.0,  0.30]},
        'legend': {'location': 'upper left'}
    },
    'contents': [
        {
            'label': r'samples ($\ell=\mu$)',
            'type': 'histogram',
            'data': {'samples': mu_samples},
            'color': 'C0'
        },
        {
            'label': r'samples ($\ell=\tau$)',
            'type': 'histogram',
            'data': {'samples': tau_samples},
            'color': 'C1'
        },
        {
            'label': r'PDF ($\ell=\mu$)',
            'type': 'signal-pdf',
            'pdf': 'B->Dlnu::dGamma/dq2;l=mu',
            'kinematic': 'q2',
            'range': [0.02, 11.60],
            'kinematics': {'q2_min':  0.02, 'q2_max': 11.60},
            'color': 'C0'
        },
        {
            'label': r'PDF ($\ell=\tau$)',
            'type': 'signal-pdf',
            'pdf': 'B->Dlnu::dGamma/dq2;l=tau',
            'kinematic': 'q2',
            'range': [3.17, 11.60],
            'kinematics': {'q2_min':  3.17, 'q2_max': 11.60},
            'color': 'C1'
        },
    ]
}
eos.plot.Plotter(plot_args).plot()
\end{lstlisting}

%--------+---------+---------+---------+---------+---------+---------+---------+
%
%
%--------+---------+---------+---------+---------+---------+---------+---------+
\section{Constraints Data Format}
\label{app:constraints-format}

Constraints are stored as \YAML~\cite{YAML} files within the \EOS source repository in the directory
\texttt{eos/constraints/}. Each constraint file is an associative array, with the
top-level keys corresponding to the constraint's qualified name, and the value describing
the constraint data. The constraint data itself is also an associative array. The
\texttt{type} key determines the type of the likelihood, and therefore which other keys
must be present. \EOS supports the following types of likelihood:
\begin{description}
    \item[\hlred{\texttt{Gaussian}}] The likelihood is a univariate Gaussian density.
    It requires the following keys:
    \smallskip
    \begin{description}
        \item[\hlgreen{\texttt{observable}}] The name of the observable that appears in this likelihood, as an \class{eos.QualifiedName}.
        \smallskip
        
        \item[\hlgreen{\texttt{kinematics}}] The kinematic variables and their values that underlay
        the likelihood's observable, as an associative array.
        \smallskip
        
        \item[\hlgreen{\texttt{options}}] The option keys and values that underlay the likelihood's
        observable, as an associative array.
        \smallskip
        
        \item[\hlgreen{\texttt{mean}}] The mean of the likelihood, as a floating point value.
        \smallskip
        
        \item[\hlgreen{\texttt{sigma-stat}}] The statistical uncertainty of the likelihood, as an associative array
        with keys \texttt{hi} and \texttt{lo}. For a completely symmetric uncertainty, set both keys to the same value.
        \smallskip
        
        \item[\hlgreen{\texttt{sigma-sys}}] The systematic uncertainty of the likelihood, as an associative array
        with keys \texttt{hi} and \texttt{lo}. For a completely symmetric uncertainty, set both keys to the same value.
        \smallskip
        
        \item[\hlgreen{\texttt{dof}}] The degrees of freedom, as a floating point value.
        Must be set to \texttt{1} to be backward compatible.
    \end{description}
    \medskip
    \item[\hlred{\texttt{MultivariateGaussian(Covariance)}}] The likelihood is a multivariate Gaussian density,
    and correlations and total uncertainties are specified through the covariance matrix.
    It requires the following keys:
    \smallskip
    \begin{description}
        \item[\hlgreen{\texttt{dim}}] The dimension of the covariance matrix, as an integer. Denoted below as $D$.
        \smallskip
        
        \item[\hlgreen{\texttt{dof}}] The degrees of freedom, as an integer.
        \smallskip
        
        \item[\hlgreen{\texttt{observables}}] The names of the observables that appear in this likelihood, as an
        ordered list of \class{eos.QualifiedName} of length $P$. Denoted below as $\vec{o}$.
        \smallskip
        
        \item[\hlgreen{\texttt{kinematics}}] The kinematic configuration for each of the observables, as an ordered list
        of length $P$ of associative arrays.
        \smallskip
        
        \item[\hlgreen{\texttt{options}}] The options for each of the observables, as an ordered list of length $P$
        of associative arrays.
        \smallskip
        
        \item[\hlgreen{\texttt{means}}] The mean values of the likelihood, as an ordered list of floating point values.
        Denoted below as $\mu$.
        \smallskip
        
        \item[\hlgreen{\texttt{covariance}}] The $D\times D$-dimensional covariance matrix of the likelihood, as an ordered list of ordered lists of floating point values (row-first ordering). Denoted below as $\Sigma$.
        \smallskip

        \item[\hlgreen{\texttt{response}}] The optional $D\times P$-dimensional response matrix that converts a $P$
        dimensional theory prediction into a $D$ dimension measurement. If not specified, \EOS assumes that $P=D$
        and that the response matrix is the identity matrix. Specified as an ordered list of ordered lists of floating point values (row-first ordering). The response matrix is used to fold the theory predictions. This enables fits involving experimental results
        that have not or cannot be unfolded. Denoted below as $R$.
        \smallskip
    \end{description}
    The logarithm of the likelihood $L$ reads
    \begin{equation}
        -2 \ln L = -2 \ln \mathcal{N}_D(R \vec{o}\,|\,\vec{\mu}, \Sigma)
        \sim \left(\vec\mu - R \vec{o}\right)^T \Sigma^{-1} \left(\vec\mu - R \vec{o}\right)\,.
    \end{equation}
    In the above, $\mathcal{N}_D(\cdot\,|\,\vec{\mu},\Sigma)$ denotes a $D$-variate Gaussian \PDF centered
    at $\mu$ with covariance $\Sigma$.
    \medskip
    \item[\hlred{\texttt{Mixture}}] The likelihood is a mixture density, with all mixture components
    being multivariate Gaussian densities. Their correlations and total uncertainties are specified
    through their respective covariance matrices.
    It requires the following keys:
    \smallskip
    \begin{description}
        \item[\hlgreen{\texttt{dim}}] The dimension of each covariance matrix, as an integer. Denoted below as $D$.
        \smallskip
        
        \item[\hlgreen{\texttt{observables}}] The names of the observables that appear in this likelihood, as an
        ordered list of \class{eos.QualifiedName} of length $D$. Denoted below as $\vec{o}$.
        \smallskip
        
        \item[\hlgreen{\texttt{kinematics}}] The kinematic configuration for each of the observables, as an ordered list
        of length $D$ of associative arrays.
        \smallskip
        
        \item[\hlgreen{\texttt{options}}] The options for each of the observables, as an ordered list of length $D$
        of associative arrays.
        \smallskip
        
        \item[\hlgreen{\texttt{components}}] The description of the mixture components as a list of length $N$.
        Each list element is an associative array that requires the following keys:
        \begin{description}
            \item[\hlorange{means}] The mean values of this component, as a list of floats of length $D$. Denoted below as
            $\vec{\mu}_n$.
            \smallskip
            \item[\hlorange{covariance}] The covariance of this component, as a list of lists of floats. Denoted below as
            $\Sigma_n$.
        \end{description}
        \smallskip
        
        \item[\hlgreen{\texttt{weights}}] The weights of the mixture components as a list of length $N$. Denoted below as $\alpha_n$.
    \end{description}

    The likelihood $L$ reads
    \begin{equation}
        L = \sum_{n=1}^N \alpha_n\, \mathcal{N}_D(\vec{o}\,|\,\vec{\mu}_n, \Sigma_n)
        \qquad \text{with} \qquad
        \sum_{n=1}^N \alpha_n = 1 \,.
    \end{equation}
\end{description}

\begin{lstlisting}[%
    language=yaml,%
    caption={
        Example of a multivariate Gaussian constraint as recorded in the \EOS source code repository,
        representing binned measurements of the $\bar{B}^0\to D^+e^-\bar\nu$ branching ratio by
        the Belle experiment~\cite{Belle:2015pkj}.
        \label{lst:constraint:Belle2015A}
    }
]
B^0->D^+e^-nu::BRs@Belle:2015A:
    type: MultivariateGaussian(Covariance)
    dim: 10
    observables:
        [ B->Dlnu::BR,    B->Dlnu::BR,    B->Dlnu::BR,    B->Dlnu::BR,    B->Dlnu::BR,
          B->Dlnu::BR,    B->Dlnu::BR,    B->Dlnu::BR,    B->Dlnu::BR,    B->Dlnu::BR  ]
    options:
        [ { l: e, q: d }, { l: e, q: d }, { l: e, q: d }, { l: e, q: d }, { l: e, q: d },
          { l: e, q: d }, { l: e, q: d }, { l: e, q: d }, { l: e, q: d }, { l: e, q: d } ]
    kinematics:
        - { q2_min: 10.44, q2_max: 11.63 }
        - { q2_min:  9.26, q2_max: 10.44 }
        - { q2_min:  8.07, q2_max:  9.26 }
        - { q2_min:  6.89, q2_max:  8.07 }
        - { q2_min:  5.71, q2_max:  6.89 }
        - { q2_min:  4.52, q2_max:  5.71 }
        - { q2_min:  3.34, q2_max:  4.52 }
        - { q2_min:  2.15, q2_max:  3.34 }
        - { q2_min:  0.97, q2_max:  2.15 }
        - { q2_min:  0.01, q2_max:  0.97 }
    means:
        [ 4.154e-05, 6.106e-04, 1.255e-03, 1.635e-03, 1.901e-03, 2.758e-03, 3.524e-03, 4.216e-03, 4.371e-03, 4.050e-03 ]
    covariance:
        - [1.912e-09, 1.726e-10, 3.427e-10, 4.424e-10, 5.041e-10, 7.320e-10, 9.624e-10, 1.096e-09, 1.049e-09, 8.840e-10]
        - [1.726e-10, 1.478e-08, 1.811e-09, 2.383e-09, 2.738e-09, 3.977e-09, 5.166e-09, 5.959e-09, 5.903e-09, 5.209e-09]
        - [3.427e-10, 1.811e-09, 2.863e-08, 4.849e-09, 5.579e-09, 8.101e-09, 1.051e-08, 1.215e-08, 1.214e-08, 1.075e-08]
        - [4.424e-10, 2.383e-09, 4.849e-09, 3.786e-08, 7.398e-09, 1.071e-08, 1.387e-08, 1.602e-08, 1.603e-08, 1.425e-08]
        - [5.041e-10, 2.738e-09, 5.579e-09, 7.398e-09, 4.355e-08, 1.241e-08, 1.606e-08, 1.860e-08, 1.873e-08, 1.678e-08]
        - [7.320e-10, 3.977e-09, 8.101e-09, 1.071e-08, 1.241e-08, 6.176e-08, 2.334e-08, 2.709e-08, 2.731e-08, 2.440e-08]
        - [9.624e-10, 5.166e-09, 1.051e-08, 1.387e-08, 1.606e-08, 2.334e-08, 8.585e-08, 3.533e-08, 3.555e-08, 3.156e-08]
        - [1.096e-09, 5.959e-09, 1.215e-08, 1.602e-08, 1.860e-08, 2.709e-08, 3.533e-08, 1.022e-07, 4.194e-08, 3.744e-08]
        - [1.049e-09, 5.903e-09, 1.214e-08, 1.603e-08, 1.873e-08, 2.731e-08, 3.555e-08, 4.194e-08, 1.005e-07, 3.887e-08]
        - [8.840e-10, 5.209e-09, 1.075e-08, 1.425e-08, 1.678e-08, 2.440e-08, 3.156e-08, 3.744e-08, 3.887e-08, 8.132e-08]
    dof: 10
\end{lstlisting}

%--------+---------+---------+---------+---------+---------+---------+---------+%
%
%
%--------+---------+---------+---------+---------+---------+---------+---------+

\printindex

%--------+---------+---------+---------+---------+---------+---------+---------+%
%
%
%--------+---------+---------+---------+---------+---------+---------+---------+
\bibliography{%
bibliography/cs.bib,%
bibliography/physics.bib,%
bibliography/statistics.bib,%
bibliography/lhcb.bib%
}

\begin{thebibliography}{100}
\providecommand{\url}[1]{{#1}}
\providecommand{\urlprefix}{URL }
\expandafter\ifx\csname urlstyle\endcsname\relax
  \providecommand{\doi}[1]{DOI \discretionary{}{}{}#1}\else
  \providecommand{\doi}{DOI \discretionary{}{}{}\begingroup
  \urlstyle{rm}\Url}\fi

\bibitem{Bevan:2014iga}
A.J. Bevan, et~al., {The Physics of the B Factories}, Eur. Phys. J. C
  \textbf{74}, 3026 (2014).
\newblock \doi{10.1140/epjc/s10052-014-3026-9}

\bibitem{LHCb:2012myk}
R.~Aaij, et~al., {Implications of LHCb measurements and future prospects}, Eur.
  Phys. J. C \textbf{73}(4), 2373 (2013).
\newblock \doi{10.1140/epjc/s10052-013-2373-2}

\bibitem{FlavourLatticeAveragingGroup:2019iem}
S.~Aoki, et~al., {FLAG Review 2019: Flavour Lattice Averaging Group (FLAG)},
  Eur. Phys. J. C \textbf{80}(2), 113 (2020).
\newblock \doi{10.1140/epjc/s10052-019-7354-7}

\bibitem{Albrecht:2021tul}
J.~Albrecht, D.~van Dyk, C.~Langenbruch, {Flavour anomalies in heavy quark
  decays}, Prog. Part. Nucl. Phys. \textbf{120}, 103885 (2021).
\newblock \doi{10.1016/j.ppnp.2021.103885}

\bibitem{Bernlochner:2021vlv}
F.U. Bernlochner, M.F. Sevilla, D.J. Robinson, G.~Wormser, {Semitauonic
  $b$-hadron decays: A lepton flavor universality laboratory},   (2021)

\bibitem{EOS}
D.~van Dyk, F.~Beaujean, T.~Blake, C.~Bobeth, M.~Bordone, E.~Eberhard,
  E.~Graverini, N.~Gubernari, A.~Kokulu, S.~Kürten, D.~Leljak, P.~Lüghausen,
  M.~Reboud, M.~Ritter, E.~Romero, I.~Toijala, K.K. Vos,
  \emph{{\textnormal{\texttt{EOS} v1.0\ --- A software for flavor physics
  phenomenology}}} (2021).
\newblock \doi{10.5281/zenodo.5730384}

\bibitem{vanDyk:2012zla}
D.~van Dyk, {The Decays $\bar{B} \to \bar{K}^{(*)} \ell^+ \ell^-$ at Low Recoil
  and their Constraints on New Physics}.
\newblock Ph.D. thesis, Dortmund U. (2012)

\bibitem{EOS:repo}
D.~van Dyk, et~al.
\newblock {\texttt{EOS} source code repository} (2021).
\newblock \url{https://github.com/eos/eos}

\bibitem{GPLv2}
Gnu general public license.
\newblock
  \urlprefix\url{https://www.gnu.org/licenses/old-licenses/gpl-2.0.html}

\bibitem{Bobeth:2010wg}
C.~Bobeth, G.~Hiller, D.~van Dyk, {The Benefits of $\bar{B} \to \bar{K}^*
  \ell^+ \ell^-$ Decays at Low Recoil}, JHEP \textbf{07}, 098 (2010).
\newblock \doi{10.1007/JHEP07(2010)098}

\bibitem{Bobeth:2011gi}
C.~Bobeth, G.~Hiller, D.~van Dyk, {More Benefits of Semileptonic Rare B Decays
  at Low Recoil: CP Violation}, JHEP \textbf{07}, 067 (2011).
\newblock \doi{10.1007/JHEP07(2011)067}

\bibitem{Bobeth:2011nj}
C.~Bobeth, G.~Hiller, D.~van Dyk, C.~Wacker, {The Decay $B \to K \ell^+ \ell^-$
  at Low Hadronic Recoil and Model-Independent $\Delta B = 1$ Constraints},
  JHEP \textbf{01}, 107 (2012).
\newblock \doi{10.1007/JHEP01(2012)107}

\bibitem{Beaujean:2012uj}
F.~Beaujean, C.~Bobeth, D.~van Dyk, C.~Wacker, {Bayesian Fit of Exclusive $b
  \to s \bar\ell\ell$ Decays: The Standard Model Operator Basis}, JHEP
  \textbf{08}, 030 (2012).
\newblock \doi{10.1007/JHEP08(2012)030}

\bibitem{Bobeth:2012vn}
C.~Bobeth, G.~Hiller, D.~van Dyk, {General analysis of $\bar{B} \to
  \bar{K}^{(*)}\ell^+ \ell^-$ decays at low recoil}, Phys. Rev. D
  \textbf{87}(3), 034016 (2013).
\newblock \doi{10.1103/PhysRevD.87.034016}

\bibitem{Beaujean:2013soa}
F.~Beaujean, C.~Bobeth, D.~van Dyk, {Comprehensive Bayesian analysis of rare
  (semi)leptonic and radiative $B$ decays}, Eur. Phys. J. C \textbf{74}, 2897
  (2014).
\newblock \doi{10.1140/epjc/s10052-014-2897-0}.
\newblock [Erratum: Eur.Phys.J.C 74, 3179 (2014)]

\bibitem{Faller:2013dwa}
S.~Faller, T.~Feldmann, A.~Khodjamirian, T.~Mannel, D.~van Dyk, {Disentangling
  the Decay Observables in $B^- \to \pi^+\pi^-\ell^-\bar\nu_\ell$}, Phys. Rev.
  D \textbf{89}(1), 014015 (2014).
\newblock \doi{10.1103/PhysRevD.89.014015}

\bibitem{SentitemsuImsong:2014plu}
I.~Sentitemsu~Imsong, A.~Khodjamirian, T.~Mannel, D.~van Dyk, {Extrapolation
  and unitarity bounds for the B \textrightarrow{} \ensuremath{\pi} form
  factor}, JHEP \textbf{02}, 126 (2015).
\newblock \doi{10.1007/JHEP02(2015)126}

\bibitem{Boer:2014kda}
P.~B\"oer, T.~Feldmann, D.~van Dyk, {Angular Analysis of the Decay $\Lambda_b
  \to \Lambda (\to N \pi) \ell^+\ell^-$}, JHEP \textbf{01}, 155 (2015).
\newblock \doi{10.1007/JHEP01(2015)155}

\bibitem{Beaujean:2015gba}
F.~Beaujean, C.~Bobeth, S.~Jahn, {Constraints on tensor and scalar couplings
  from $B\rightarrow K\bar{\mu }\mu $ and $B_s\rightarrow \bar{\mu }\mu $},
  Eur. Phys. J. C \textbf{75}(9), 456 (2015).
\newblock \doi{10.1140/epjc/s10052-015-3676-2}

\bibitem{Feldmann:2015xsa}
T.~Feldmann, B.~M\"uller, D.~van Dyk, {Analyzing $b\to u$ transitions in
  semileptonic $\bar{B}_s \to K^{*+}(\to K \pi)\ell^-\bar\nu_\ell$ decays},
  Phys. Rev. D \textbf{92}(3), 034013 (2015).
\newblock \doi{10.1103/PhysRevD.92.034013}

\bibitem{Mannel:2015osa}
T.~Mannel, D.~van Dyk, {Zero-recoil sum rules for $\Lambda_b \to \Lambda_c$
  form factors}, Phys. Lett. B \textbf{751}, 48 (2015).
\newblock \doi{10.1016/j.physletb.2015.10.016}

\bibitem{Bordone:2016tex}
M.~Bordone, G.~Isidori, D.~van Dyk, {Impact of leptonic $\tau $ decays on the
  distribution of $B\rightarrow P\mu \bar{\nu }$ decays}, Eur. Phys. J. C
  \textbf{76}(7), 360 (2016).
\newblock \doi{10.1140/epjc/s10052-016-4202-x}

\bibitem{Meinel:2016grj}
S.~Meinel, D.~van Dyk, {Using $\Lambda_b\to \Lambda\mu^+\mu^-$ data within a
  Bayesian analysis of $|\Delta B| = |\Delta S| = 1$ decays}, Phys. Rev. D
  \textbf{94}(1), 013007 (2016).
\newblock \doi{10.1103/PhysRevD.94.013007}

\bibitem{Boer:2016iez}
P.~B\"oer, T.~Feldmann, D.~van Dyk, {QCD Factorization Theorem for $B \to
  \pi\pi\ell\nu$ Decays at Large Dipion Masses}, JHEP \textbf{02}, 133 (2017).
\newblock \doi{10.1007/JHEP02(2017)133}

\bibitem{Serra:2016ivr}
N.~Serra, R.~Silva~Coutinho, D.~van Dyk, {Measuring the breaking of lepton
  flavor universality in $B\to K^*\ell^+\ell^-$}, Phys. Rev. D \textbf{95}(3),
  035029 (2017).
\newblock \doi{10.1103/PhysRevD.95.035029}

\bibitem{Bobeth:2017vxj}
C.~Bobeth, M.~Chrzaszcz, D.~van Dyk, J.~Virto, {Long-distance effects in
  $B\rightarrow K^*\ell \ell $ from analyticity}, Eur. Phys. J. C
  \textbf{78}(6), 451 (2018).
\newblock \doi{10.1140/epjc/s10052-018-5918-6}

\bibitem{Blake:2017une}
T.~Blake, M.~Kreps, {Angular distribution of polarised $\Lambda_b$ baryons
  decaying to $\Lambda \ell^+\ell^-$}, JHEP \textbf{11}, 138 (2017).
\newblock \doi{10.1007/JHEP11(2017)138}

\bibitem{Boer:2018vpx}
P.~B\"oer, M.~Bordone, E.~Graverini, P.~Owen, M.~Rotondo, D.~van Dyk, {Testing
  lepton flavour universality in semileptonic $\Lambda_b \to \Lambda_c^*$
  decays}, JHEP \textbf{06}, 155 (2018).
\newblock \doi{10.1007/JHEP06(2018)155}

\bibitem{Feldmann:2018kqr}
T.~Feldmann, D.~van Dyk, K.K. Vos, {Revisiting $B \to \pi\pi \ell \nu$ at large
  dipion masses}, JHEP \textbf{10}, 030 (2018).
\newblock \doi{10.1007/JHEP10(2018)030}

\bibitem{Gubernari:2018wyi}
N.~Gubernari, A.~Kokulu, D.~van Dyk, {$B\to P$ and $B\to V$ Form Factors from
  $B$-Meson Light-Cone Sum Rules beyond Leading Twist}, JHEP \textbf{01}, 150
  (2019).
\newblock \doi{10.1007/JHEP01(2019)150}

\bibitem{Boer:2019zmp}
P.~B\"oer, A.~Kokulu, J.N. Toelstede, D.~van Dyk, {Angular Analysis of
  \textbackslash{}boldmath $\Lambda_b\to \Lambda_c (\to \Lambda
  \pi)\ell\bar\nu$}, JHEP \textbf{12}, 082 (2019).
\newblock \doi{10.1007/JHEP12(2019)082}

\bibitem{Bordone:2019vic}
M.~Bordone, M.~Jung, D.~van Dyk, {Theory determination of $\bar{B}\to
  D^{(*)}\ell^-\bar\nu$ form factors at $\mathcal{O}(1/m_c^2)$}, Eur. Phys. J.
  C \textbf{80}(2), 74 (2020).
\newblock \doi{10.1140/epjc/s10052-020-7616-4}

\bibitem{Blake:2019guk}
T.~Blake, S.~Meinel, D.~van Dyk, {Bayesian Analysis of $b\to s\mu^+\mu^-$
  Wilson Coefficients using the Full Angular Distribution of $\Lambda_b\to
  \Lambda(\to p\, \pi^-)\mu^+\mu^-$ Decays}, Phys. Rev. D \textbf{101}(3),
  035023 (2020).
\newblock \doi{10.1103/PhysRevD.101.035023}

\bibitem{Bordone:2019guc}
M.~Bordone, N.~Gubernari, D.~van Dyk, M.~Jung, {Heavy-Quark expansion for
  ${{\bar{B}}_s\rightarrow D^{(*)}_s}$ form factors and unitarity bounds beyond
  the ${SU(3)_F}$ limit}, Eur. Phys. J. C \textbf{80}(4), 347 (2020).
\newblock \doi{10.1140/epjc/s10052-020-7850-9}

\bibitem{Gubernari:2020eft}
N.~Gubernari, D.~van Dyk, J.~Virto, {Non-local matrix elements in $B_{(s)}\to
  \{K^{(*)},\phi\}\ell^+\ell^-$}, JHEP \textbf{02}, 088 (2021).
\newblock \doi{10.1007/JHEP02(2021)088}

\bibitem{Bruggisser:2021duo}
S.~Bruggisser, R.~Sch\"afer, D.~van Dyk, S.~Westhoff, {The Flavor of UV
  Physics}, JHEP \textbf{05}, 257 (2021).
\newblock \doi{10.1007/JHEP05(2021)257}

\bibitem{Leljak:2021vte}
D.~Leljak, B.~Meli\'c, D.~van Dyk, {The $\bar{B} \to \pi$ form factors from QCD
  and their impact on $|V_{ub}|$}, JHEP \textbf{07}, 036 (2021).
\newblock \doi{10.1007/JHEP07(2021)036}

\bibitem{Bobeth:2021lya}
C.~Bobeth, M.~Bordone, N.~Gubernari, M.~Jung, D.~van Dyk, {Lepton-flavour
  non-universality of ${\bar{B}}\rightarrow D^*\ell {{\bar{\nu }}}$ angular
  distributions in and beyond the Standard Model}, Eur. Phys. J. C
  \textbf{81}(11), 984 (2021).
\newblock \doi{10.1140/epjc/s10052-021-09724-2}

\bibitem{CDF:2011tds}
T.~Aaltonen, et~al., {Measurements of the Angular Distributions in the Decays
  $B \to K^{(*)} \mu^+ \mu^-$ at CDF}, Phys. Rev. Lett. \textbf{108}, 081807
  (2012).
\newblock \doi{10.1103/PhysRevLett.108.081807}

\bibitem{CMS:2013mkz}
S.~Chatrchyan, et~al., {Angular Analysis and Branching Fraction Measurement of
  the Decay $B^0 \to K^{*0} \mu^+\mu^-$}, Phys. Lett. B \textbf{727}, 77
  (2013).
\newblock \doi{10.1016/j.physletb.2013.10.017}

\bibitem{CMS:2015bcy}
V.~Khachatryan, et~al., {Angular analysis of the decay $B^0 \to K^{*0} \mu^+
  \mu^-$ from pp collisions at $\sqrt s = 8$ TeV}, Phys. Lett. B \textbf{753},
  424 (2016).
\newblock \doi{10.1016/j.physletb.2015.12.020}

\bibitem{LHCb:2012bin}
R.~Aaij, et~al., {Measurement of the isospin asymmetry in $B \to
  K^{(*)}\mu^+\mu^-$ decays}, JHEP \textbf{07}, 133 (2012).
\newblock \doi{10.1007/JHEP07(2012)133}

\bibitem{LHCb:2013zuf}
R.~Aaij, et~al., {Differential branching fraction and angular analysis of the
  decay $B^{0} \to K^{*0} \mu^{+}\mu^{-}$}, JHEP \textbf{08}, 131 (2013).
\newblock \doi{10.1007/JHEP08(2013)131}

\bibitem{LHCb:2014auh}
R.~Aaij, et~al., {Angular analysis of charged and neutral $B \to K \mu^+\mu^-$
  decays}, JHEP \textbf{05}, 082 (2014).
\newblock \doi{10.1007/JHEP05(2014)082}

\bibitem{LHCb:2015svh}
R.~Aaij, et~al., {Angular analysis of the $B^{0} \to K^{*0} \mu^{+} \mu^{-}$
  decay using 3 fb$^{-1}$ of integrated luminosity}, JHEP \textbf{02}, 104
  (2016).
\newblock \doi{10.1007/JHEP02(2016)104}

\bibitem{LHCb:2018jna}
R.~Aaij, et~al., {Angular moments of the decay $\Lambda_b^0 \rightarrow \Lambda
  \mu^{+} \mu^{-}$ at low hadronic recoil}, JHEP \textbf{09}, 146 (2018).
\newblock \doi{10.1007/JHEP09(2018)146}

\bibitem{LHCb:2020lmf}
R.~Aaij, et~al., {Measurement of $CP$-Averaged Observables in the
  $B^{0}\rightarrow K^{*0}\mu^{+}\mu^{-}$ Decay}, Phys. Rev. Lett.
  \textbf{125}(1), 011802 (2020).
\newblock \doi{10.1103/PhysRevLett.125.011802}

\bibitem{basf2ext}
{Belle-II Framework Software Group}.
\newblock \url{https://github.com/belle2/externals}

\bibitem{Kuhr:2018lps}
T.~Kuhr, C.~Pulvermacher, M.~Ritter, T.~Hauth, N.~Braun, {The Belle II Core
  Software}, Comput. Softw. Big Sci. \textbf{3}(1), 1 (2019).
\newblock \doi{10.1007/s41781-018-0017-9}

\bibitem{python}
G.~Van~Rossum, F.L. Drake~Jr, \emph{Python tutorial} (Centrum voor Wiskunde en
  Informatica Amsterdam, The Netherlands, 1995)

\bibitem{Harris:2020xlr}
C.R. Harris, et~al., {Array programming with NumPy}, Nature \textbf{585}(7825),
  357 (2020).
\newblock \doi{10.1038/s41586-020-2649-2}

\bibitem{pypmc}
F.~Beaujean, S.~Jahn.
\newblock {\texttt{pypmc} online documentation} (2021).
\newblock \url{https://pypmc.github.io}

\bibitem{jupyter}
T.~Kluyver, B.~Ragan-Kelley, F.~P{\'e}rez, B.~Granger, M.~Bussonnier,
  J.~Frederic, K.~Kelley, J.~Hamrick, J.~Grout, S.~Corlay, P.~Ivanov, D.~Avila,
  S.~Abdalla, C.~Willing, {Jupyter development team}, in \emph{Positioning and
  Power in Academic Publishing: Players, Agents and Agendas}, ed. by
  F.~Loizides, B.~Schmidt (IOS Press, Netherlands, 2016), pp. 87--90.
\newblock \urlprefix\url{https://eprints.soton.ac.uk/403913/}

\bibitem{PEP-600}
N.J. Smith, T.~Kluyver.
\newblock {Future 'manylinux' Platform Tags for Portable Linux Built
  Distributions} (2019).
\newblock \url{https://www.python.org/dev/peps/pep-0600/}

\bibitem{EOS:doc}
D.~van Dyk, et~al.
\newblock {\texttt{EOS} v1.0 online documentation} (2021).
\newblock \url{https://eos.github.io/doc/v1.0/}

\bibitem{Straub:2018kue}
D.M. Straub, {flavio: a Python package for flavour and precision phenomenology
  in the Standard Model and beyond},   (2018)

\bibitem{Neshatpour:2021nbn}
S.~Neshatpour, F.~Mahmoudi, {Flavour Physics with SuperIso}, PoS
  \textbf{TOOLS2020}, 036 (2021).
\newblock \doi{10.22323/1.392.0036}

\bibitem{DeBlas:2019ehy}
J.~De~Blas, et~al., {$\texttt{HEPfit}$: a code for the combination of indirect
  and direct constraints on high energy physics models}, Eur. Phys. J. C
  \textbf{80}(5), 456 (2020).
\newblock \doi{10.1140/epjc/s10052-020-7904-z}

\bibitem{Workgroup:2017myk}
F.U. Bernlochner, et~al., {FlavBit: A GAMBIT module for computing flavour
  observables and likelihoods}, Eur. Phys. J. C \textbf{77}(11), 786 (2017).
\newblock \doi{10.1140/epjc/s10052-017-5157-2}

\bibitem{bsll2021}
N.~Gubernari, M.~Reboud, D.~van Dyk, J.~Virto, ,   (2021).
\newblock To appear

\bibitem{EOS:API}
D.~van Dyk, et~al.
\newblock {\texttt{EOS} v1.0 Python API} (2021).
\newblock \url{https://eos.github.io/doc/v1.0/api/python.html}

\bibitem{EOS:examples}
D.~van Dyk, et~al.
\newblock {\texttt{EOS} example notebooks} (2021).
\newblock \url{https://github.com/eos/eos/tree/v1.0/examples/}

\bibitem{Straub:2015ica}
A.~Bharucha, D.M. Straub, R.~Zwicky, {$B\to V\ell^+\ell^-$ in the Standard
  Model from light-cone sum rules}, JHEP \textbf{08}, 098 (2016).
\newblock \doi{10.1007/JHEP08(2016)098}

\bibitem{crooks2015amoroso}
G.E. Crooks, The amoroso distribution,   (2015)

\bibitem{Lattice:2015rga}
J.A. Bailey, et~al., {$B\to D \ell \nu$ form factors at nonzero recoil and
  $|V_{cb}|$ from 2+1-flavor lattice QCD}, Phys. Rev. D \textbf{92}(3), 034506
  (2015).
\newblock \doi{10.1103/PhysRevD.92.034506}

\bibitem{Belle:2015pkj}
R.~Glattauer, et~al., {Measurement of the decay $B\to D\ell\nu_\ell$ in fully
  reconstructed events and determination of the Cabibbo-Kobayashi-Maskawa
  matrix element $|V_{cb}|$}, Phys. Rev. D \textbf{93}(3), 032006 (2016).
\newblock \doi{10.1103/PhysRevD.93.032006}

\bibitem{Na:2015kha}
H.~Na, C.M. Bouchard, G.P. Lepage, C.~Monahan, J.~Shigemitsu, {$B \rightarrow D
  \ell \nu$ form factors at nonzero recoil and extraction of $|V_{cb}|$}, Phys.
  Rev. D \textbf{92}(5), 054510 (2015).
\newblock \doi{10.1103/PhysRevD.93.119906}.
\newblock [Erratum: Phys.Rev.D 93, 119906 (2016)]

\bibitem{MILC:2015uhg}
J.A. Bailey, et~al., {$B\to D\ell\nu$ form factors at nonzero recoil and
  $|V_{cb}|$ from 2+1-flavor lattice QCD}, Phys. Rev. D \textbf{92}(3), 034506
  (2015).
\newblock \doi{10.1103/PhysRevD.92.034506}

\bibitem{doi:10.1063/1.1699114}
N.~Metropolis, A.W. Rosenbluth, M.N. Rosenbluth, A.H. Teller, E.~Teller,
  Equation of state calculations by fast computing machines, The Journal of
  Chemical Physics \textbf{21}(6), 1087 (1953).
\newblock \doi{10.1063/1.1699114}.
\newblock \urlprefix\url{https://doi.org/10.1063/1.1699114}

\bibitem{10.1093/biomet/57.1.97}
W.K. Hastings, {Monte Carlo sampling methods using Markov chains and their
  applications}, Biometrika \textbf{57}(1), 97 (1970).
\newblock \doi{10.1093/biomet/57.1.97}.
\newblock \urlprefix\url{https://doi.org/10.1093/biomet/57.1.97}

\bibitem{10.2307/3318737}
H.~Haario, E.~Saksman, J.~Tamminen, An adaptive metropolis algorithm, Bernoulli
  \textbf{7}(2), 223 (2001).
\newblock \urlprefix\url{http://www.jstor.org/stable/3318737}

\bibitem{2010MNRAS.405.2381K}
M.~{Kilbinger}, D.~{Wraith}, C.P. {Robert}, K.~{Benabed}, O.~{Capp{\'e}}, J.F.
  {Cardoso}, G.~{Fort}, S.~{Prunet}, F.R. {Bouchet}, {Bayesian model comparison
  in cosmology with Population Monte Carlo}, MNRAS \textbf{405}(4), 2381
  (2010).
\newblock \doi{10.1111/j.1365-2966.2010.16605.x}

\bibitem{2013arXiv1304.7808B}
F.~{Beaujean}, A.~{Caldwell}, {Initializing adaptive importance sampling with
  Markov chains},  arXiv:1304.7808 (2013)

\bibitem{matplotlib}
J.D. Hunter, Matplotlib: A 2d graphics environment, Computing in Science \&
  Engineering \textbf{9}(3), 90 (2007).
\newblock \doi{10.1109/MCSE.2007.55}

\bibitem{gelmanbda04}
A.~Gelman, J.B. Carlin, H.S. Stern, D.B. Rubin, \emph{Bayesian Data Analysis},
  2nd edn. (Chapman and Hall/CRC, 2004)

\bibitem{scipy}
P.~Virtanen, R.~Gommers, T.E. Oliphant, M.~Haberland, T.~Reddy, D.~Cournapeau,
  E.~Burovski, P.~Peterson, W.~Weckesser, J.~Bright, S.J. {van der Walt},
  M.~Brett, J.~Wilson, K.J. Millman, N.~Mayorov, A.R.J. Nelson, E.~Jones,
  R.~Kern, E.~Larson, C.J. Carey, {\.I}.~Polat, Y.~Feng, E.W. Moore,
  J.~{VanderPlas}, D.~Laxalde, J.~Perktold, R.~Cimrman, I.~Henriksen, E.A.
  Quintero, C.R. Harris, A.M. Archibald, A.H. Ribeiro, F.~Pedregosa, P.~{van
  Mulbregt}, {SciPy 1.0 Contributors}, {{SciPy} 1.0: Fundamental Algorithms for
  Scientific Computing in Python}, Nature Methods \textbf{17}, 261 (2020).
\newblock \doi{10.1038/s41592-019-0686-2}

\bibitem{Wolfenstein:1983yz}
L.~Wolfenstein, {Parametrization of the Kobayashi-Maskawa Matrix}, Phys. Rev.
  Lett. \textbf{51}, 1945 (1983).
\newblock \doi{10.1103/PhysRevLett.51.1945}

\bibitem{Charles:2004jd}
J.~Charles, A.~Hocker, H.~Lacker, S.~Laplace, F.R. Le~Diberder, J.~Malcles,
  J.~Ocariz, M.~Pivk, L.~Roos, {CP violation and the CKM matrix: Assessing the
  impact of the asymmetric $B$ factories}, Eur. Phys. J. C \textbf{41}(1), 1
  (2005).
\newblock \doi{10.1140/epjc/s2005-02169-1}

\bibitem{Buchalla:1995vs}
G.~Buchalla, A.J. Buras, M.E. Lautenbacher, {Weak decays beyond leading
  logarithms}, Rev. Mod. Phys. \textbf{68}, 1125 (1996).
\newblock \doi{10.1103/RevModPhys.68.1125}

\bibitem{Buras:1998raa}
A.J. Buras, in \emph{{Les Houches Summer School in Theoretical Physics, Session
  68: Probing the Standard Model of Particle Interactions}} (1998)

\bibitem{Buras:2011we}
A.J. Buras, {Climbing NLO and NNLO Summits of Weak Decays},   (2011)

\bibitem{Aebischer:2017gaw}
J.~Aebischer, M.~Fael, C.~Greub, J.~Virto, {B physics Beyond the Standard Model
  at One Loop: Complete Renormalization Group Evolution below the Electroweak
  Scale}, JHEP \textbf{09}, 158 (2017).
\newblock \doi{10.1007/JHEP09(2017)158}

\bibitem{Jenkins:2017jig}
E.E. Jenkins, A.V. Manohar, P.~Stoffer, {Low-Energy Effective Field Theory
  below the Electroweak Scale: Operators and Matching}, JHEP \textbf{03}, 016
  (2018).
\newblock \doi{10.1007/JHEP03(2018)016}

\bibitem{Aebischer:2017ugx}
J.~Aebischer, et~al., {WCxf: an exchange format for Wilson coefficients beyond
  the Standard Model}, Comput. Phys. Commun. \textbf{232}, 71 (2018).
\newblock \doi{10.1016/j.cpc.2018.05.022}

\bibitem{wcxf:EOS-basis}
{The WCxf Authors}.
\newblock {\texttt{EOS} WET basis} (2021).
\newblock \url{https://wcxf.github.io/assets/pdf/WET.EOS.pdf}

\bibitem{Aebischer:2018bkb}
J.~Aebischer, J.~Kumar, D.M. Straub, {Wilson: a Python package for the running
  and matching of Wilson coefficients above and below the electroweak scale},
  Eur. Phys. J. C \textbf{78}(12), 1026 (2018).
\newblock \doi{10.1140/epjc/s10052-018-6492-7}

\bibitem{Buchmuller:1985jz}
W.~Buchmuller, D.~Wyler, {Effective Lagrangian Analysis of New Interactions and
  Flavor Conservation}, Nucl. Phys. B \textbf{268}, 621 (1986).
\newblock \doi{10.1016/0550-3213(86)90262-2}

\bibitem{Grzadkowski:2010es}
B.~Grzadkowski, M.~Iskrzynski, M.~Misiak, J.~Rosiek, {Dimension-Six Terms in
  the Standard Model Lagrangian}, JHEP \textbf{10}, 085 (2010).
\newblock \doi{10.1007/JHEP10(2010)085}

\bibitem{Adel:1993ah}
K.~Adel, Y.P. Yao, {$O(\alpha_s)$ calculation of the decays $b \to s \gamma$
  and $b \to s g$}, Phys. Rev. D \textbf{49}, 4945 (1994).
\newblock \doi{10.1103/PhysRevD.49.4945}

\bibitem{Greub:1997hf}
C.~Greub, T.~Hurth, {Two loop matching of the dipole operators for $b \to s
  \gamma$ and $b \to s g$}, Phys. Rev. D \textbf{56}, 2934 (1997).
\newblock \doi{10.1103/PhysRevD.56.2934}

\bibitem{Bobeth:1999mk}
C.~Bobeth, M.~Misiak, J.~Urban, {Photonic penguins at two loops and $m_t$
  dependence of $BR[B \to X_s \ell^+ \ell^-]$}, Nucl. Phys. B \textbf{574}, 291
  (2000).
\newblock \doi{10.1016/S0550-3213(00)00007-9}

\bibitem{Chetyrkin:1996vx}
K.G. Chetyrkin, M.~Misiak, M.~M{\"u}nz, {Weak radiative $B$ meson decay beyond
  leading logarithms}, Phys. Lett. B \textbf{400}, 206 (1997).
\newblock \doi{10.1016/S0370-2693(97)00324-9}.
\newblock [Erratum: Phys.Lett.B 425, 414 (1998)]

\bibitem{Bobeth:2003at}
C.~Bobeth, P.~Gambino, M.~Gorbahn, U.~Haisch, {Complete NNLO QCD analysis of $B
  \to X_s \ell^+ \ell^-$ and higher order electroweak effects}, JHEP
  \textbf{04}, 071 (2004).
\newblock \doi{10.1088/1126-6708/2004/04/071}

\bibitem{Gorbahn:2004my}
M.~Gorbahn, U.~Haisch, {Effective Hamiltonian for non-leptonic $|\Delta F| = 1$
  decays at NNLO in QCD}, Nucl. Phys. B \textbf{713}, 291 (2005).
\newblock \doi{10.1016/j.nuclphysb.2005.01.047}

\bibitem{Gorbahn:2005sa}
M.~Gorbahn, U.~Haisch, M.~Misiak, {Three-loop mixing of dipole operators},
  Phys. Rev. Lett. \textbf{95}, 102004 (2005).
\newblock \doi{10.1103/PhysRevLett.95.102004}

\bibitem{Huber:2005ig}
T.~Huber, E.~Lunghi, M.~Misiak, D.~Wyler, {Electromagnetic logarithms in $B \to
  X_s \ell^+ \ell^-$}, Nucl. Phys. B \textbf{740}, 105 (2006).
\newblock \doi{10.1016/j.nuclphysb.2006.01.037}

\bibitem{Misiak:1999yg}
M.~Misiak, J.~Urban, {QCD corrections to FCNC decays mediated by Z penguins and
  W boxes}, Phys. Lett. B \textbf{451}, 161 (1999).
\newblock \doi{10.1016/S0370-2693(99)00150-1}

\bibitem{Buchalla:1998ba}
G.~Buchalla, A.J. Buras, {The rare decays $K\to \pi \nu\bar\nu$, $B \to X
  \nu\bar\nu$ and $B \to l^+ l^-$: An Update}, Nucl. Phys. B \textbf{548}, 309
  (1999).
\newblock \doi{10.1016/S0550-3213(99)00149-2}

\bibitem{Sirlin:1981ie}
A.~Sirlin, {Large $m_W, m_Z$ Behavior of the $O(\alpha)$ Corrections to
  Semileptonic Processes Mediated by W}, Nucl. Phys. B \textbf{196}, 83 (1982).
\newblock \doi{10.1016/0550-3213(82)90303-0}

\bibitem{Buras:1990fn}
A.J. Buras, M.~Jamin, P.H. Weisz, {Leading and Next-to-leading {QCD}
  Corrections to $\epsilon$ Parameter and $B^0 - \bar{B}^0$ Mixing in the
  Presence of a Heavy Top Quark}, Nucl. Phys. B \textbf{347}, 491 (1990).
\newblock \doi{10.1016/0550-3213(90)90373-L}

\bibitem{Gambino:1998rt}
P.~Gambino, A.~Kwiatkowski, N.~Pott, {Electroweak effects in the $B^0 -
  \bar{B}^0$ mixing}, Nucl. Phys. B \textbf{544}, 532 (1999).
\newblock \doi{10.1016/S0550-3213(98)00860-8}

\bibitem{Buras:2000if}
A.J. Buras, M.~Misiak, J.~Urban, {Two loop QCD anomalous dimensions of flavor
  changing four quark operators within and beyond the standard model}, Nucl.
  Phys. B \textbf{586}, 397 (2000).
\newblock \doi{10.1016/S0550-3213(00)00437-5}

\bibitem{YAML}
O.~Ben-Kiki, C.~Evans, I.~d{\"o}t Net.
\newblock {YAML Ain’t Markup Language (YAML\texttrademark) version 1.2}
  (2021).
\newblock \url{https://yaml.org/spec/1.2.2/}

\end{thebibliography}
\bibliographystyle{bibliography/spphys}

\end{document}